\documentclass[%
 reprint,
 amsmath,amssymb,
 aps,
showkeys
]{revtex4-1}

\usepackage{graphicx}
\usepackage{dcolumn}
\usepackage{bm}

\usepackage{xcolor}

\begin{document}


\title{Turbulent Prandtl number and characteristic length scales in stably stratified flows: steady-state analytical solutions}

\author{Sukanta Basu}
\email{sukanta.basu@gmail.com}
\affiliation{Faculty of Civil Engineering and Geosciences, Delft University of Technology, Delft, the Netherlands
}
\author{Albert A. M. Holtslag}
\email{bert.holtslag@wur.nl}
\affiliation{Meteorology and Air Quality, Wageningen University, Wageningen, the Netherlands}%

\date{\today}

\begin{abstract}
In this study, the stability dependence of turbulent Prandtl number ($Pr_t$) is quantified via a novel and simple analytical approach. Based on the variance and flux budget equations, a hybrid length scale formulation is first proposed and its functional relationships to well-known length scales are established. Next, the ratios of these length scales are utilized to derive an explicit relationship between $Pr_t$ and gradient Richardson number. In addition, theoretical predictions are made for several key turbulence variables (e.g., dissipation rates, normalized fluxes). The results from our proposed approach are compared against other competing formulations as well as published datasets. Overall, the agreement between the different approaches is rather good despite their different theoretical foundations and assumptions. 
\end{abstract}

\keywords{Anisotropy; Buoyancy length scale; Gradient Richardson number; Shear length scale; Stable boundary layer}

\maketitle

\section{Introduction}
\label{intro}

According to the K-theory, based on the celebrated hypothesis of Boussinesq in 1877, turbulent fluxes can be approximated as products of the eddy exchange coefficients (known as the Austausch coefficients in earlier literature) and the mean gradients  \cite{lumley64}. Specifically, for incompressible, horizontally homogeneous, boundary layer flows, the along-wind momentum flux ($\overline{u'w'}$) and the sensible heat flux ($\overline{w'\theta'}$) can be simply written as follows: 

\begin{subequations}
\begin{equation}
    \overline{u'w'} = -K_M S,
    \label{EqK1}
\end{equation}
\begin{equation}
    \overline{w'\theta'} = -K_H \Gamma.
    \label{EqK2}
\end{equation}
\end{subequations}
Here $S$ and $\Gamma$ denote the vertical gradients of the mean along-wind velocity component and the mean potential temperature, respectively. The eddy viscosity and diffusivity for heat are represented by $K_M$ and $K_H$, respectively. In contrast to molecular diffusivities, these eddy exchange coefficients are not intrinsic properties of the fluid \cite{arya01,stull88}; rather, they depend on the nature of the turbulent flows (e.g., stability) and position in the flow (e.g., distance from the wall).  

The ratio of $K_M$ and $K_H$ is known as the turbulent Prandtl number: 
\begin{equation}
    Pr_t = \frac{K_M}{K_H}.
    \label{EqPr}
\end{equation}
This variable is fundamentally different from the molecular Prandtl number: 
\begin{equation}
    Pr_m = \frac{\nu}{\alpha},
\end{equation}
where, $\nu$ and $\alpha$ denote kinematic viscosity and thermal diffusivity, respectively. According to a vast amount of literature, $Pr_t$ is strongly dependent on buoyancy and somewhat weakly dependent on other factors (see below). 

For non-buoyant (also called neutral) flows, in this paper, the turbulent Prandtl number is denoted as $Pr_{t0}$. In the past, for simplicity, a number of studies assumed $Pr_{t0} = 1$ by invoking the so-called `Reynolds analogy' hypothesis \cite{reynolds74,sutton55,tennekes72}. Basically, they implicitly assume that the turbulent transport of momentum and heat are identical. However, this assumption of $Pr_{t0} = 1$ is not supported by the vast majority of experimental data (see \cite{kays94} and the references therein). On this issue, Launder \cite{launder78} commented:
\begin{quotation}
``It would also be helpful to dispel the idea that a turbulent Prandtl number of unity was in any sense the ``normal'' value. We shall see [...] that a value of about 0.7 has a far stronger claim to normality.''
\end{quotation}
Perhaps, it is not a mere coincidence that the theoretical study of Yakhot et al. \cite{yakhot87} predicted that $Pr_{t0}$ asymptotically approaches 0.7179 in the limit of infinite $Re$ (see also \cite{sukoriansky05}). One of the most cited studies in atmospheric science, by Businger et al. \cite{businger71}, also reported $Pr_{t0} =$ 0.74. According to a review article by Kays \cite{kays94}, for laboratory flows, $Pr_{t0}$ typically falls within the range of 0.7 to 0.9; the most frequent value being equal to 0.85. Most commercial computational fluid dynamics packages (e.g., Fluent, OpenFOAM) assume 0.85 to be the default $Pr_{t0}$ value. 

There is some evidence that $Pr_{t0}$ may not be a universal constant; it might weakly depend on $Pr_m$, $Re$, and/or position in the flow. However, there is no general agreement in the literature on this matter (e.g., \cite{antonia91}). Reynolds \cite{reynolds75} summarized numerous empirical and semi-empirical formulations capturing such dependencies for a wide range of fluids (including air, water, liquid metal) and engineering flows (e.g., pipe flow, jet flow, shear flow). However, to the best of our knowledge, these formulations are yet to be confirmed for high-$Re$ atmospheric flows. In such flows, buoyancy effects have been found to be far more dominant than any other factors. 

In atmospheric flows, especially under stably stratified conditions, the value of $Pr_t$ departs significantly from $Pr_{t0}$. Over the decades, several empirical formulations have been developed by various research groups (see \cite{li19} for a recent review). For example, by regression analysis of aircraft measurements from different field campaigns, Kim and Mahrt \cite{kim92} proposed: 
\begin{equation}
    Pr_t = 1 + 3.8 Ri_g, 
\end{equation}
where, $Ri_g$ is the gradient Richardson number, commonly used to quantify atmospheric stability. It is defined as follows: 
\begin{equation}
    Ri_g = \frac{\left(\frac{g}{\Theta_0}\right)\Gamma}{S^2} = \frac{\beta \Gamma}{S^2} = \frac{N^2}{S^2}.
\end{equation}
Where,  $g$ is the gravitational acceleration and $\Theta_0$ represents a reference temperature. The variable $\beta$ is known as the buoyancy parameter. The so-called Brunt V\"ais\"al\"a frequency is denoted by $N$. 

More recently, Anderson~\cite{anderson09} conducted rigorous statistical analysis of observational data from the Antarctic. By avoiding the self-correlation issue, he proposed the following empirical relationship for $0.01 < Ri_g < 0.25$:
\begin{equation}
    Pr_t^{-1} = (0.84 \pm 0.03) Ri_g^{-0.105\pm0.012}.
\end{equation}
Clearly, the $Ri_g$-dependence of $Pr_t$ becomes rather weak as the stability of the flow decreases. 

In addition to field observational data, laboratory and simulated data were also utilized to quantify the $Pr_t$--$Ri_g$ relationship. In this regard, a popular semi-empirical formulation by Schumann and Gerz \cite{schumann95} is worth noting: 
\begin{equation}
    Pr_t = Pr_{t0} \exp\left(- \frac{Ri_g}{Pr_{t0}R_{f\infty}}\right) + \frac{Ri_g}{R_{f\infty}},
    \label{SG95}
\end{equation}
where, $R_{f\infty}$ is the asymptotic value of the flux Richardson number ($R_f = Ri_g/Pr_t$) for strongly stratified conditions. Recently, Venayagamoorthy and Stretch \cite{venayagamoorthy10} used direct numerical simulation (DNS) data and revised the formulation by Eq.~(\ref{SG95}) as follows: 
\begin{equation}
    Pr_t = Pr_{t0} \exp\left[- \frac{Ri_g\left(1 - R_{f\infty} \right)}{Pr_{t0}R_{f\infty}}\right] + \frac{Ri_g}{R_{f\infty}}. 
    \label{VS10}
\end{equation}
For all practical purposes, the differences between Eq.~(\ref{SG95}) and Eq.~(\ref{VS10}) are quite small. 

In parallel to observational and simulation studies, there have been a handful of attempts to derive the $Pr_t$\textemdash$Ri_g$ formulations from the governing equations with certain assumptions. In the appendices, we have summarized two competing hypotheses by Katul et al.~\cite{katul14} and Zilitinkevich et al.~\cite{zilitinkevich13}. The readers are also encouraged to peruse the following papers describing other relevant hypotheses: \cite{cheng02}, \cite{cheng20}, and \cite{kantha18}. In the present study, we report an alternative analytical derivation which leads to a closed-form $Pr_t$\textemdash$Ri_g$ relationship.    

\section{Analytical Derivations}
\label{Derivations}

In this section, based on the variance and flux budget equations, we first derive a hybrid length scale ($L_X$) and establish its relationship with three well-known length scales: the Hunt length scale ($L_H$, \cite{hunt89,hunt88}), the buoyancy length scale ($L_b$, \cite{brost78,wyngaard10}), and the Ellison length scale ($L_E$, \cite{ellison57}). Next, the ratios of various length scales (e.g., $L_b/L_E$) are shown to be explicit functions of $Ri_g$ and $Pr_t$. Equating these functions with one another results in a quadratic equation for $Pr_t$. One of the roots of this quadratic equation provides an explicit $Pr_t$\textemdash$Ri_g$ relationship. 

\subsection{Budget Equations}

The simplified budget equations for turbulent kinetic energy (TKE), variance of temperature ($\sigma_\theta^2$), and sensible heat flux ($\overline{w'\theta'}$) can be written as \cite{fitzjarrald79,nieuwstadt84,wyngaard75}:

\begin{subequations}
\begin{align}
    \overline{\varepsilon} & = - \left(\overline{u'w'}\right) S + \beta \overline{w'\theta'},
    \label{EqEDR1}\\
    \overline{\chi}_{\theta} & = -2 \left(\overline{w'\theta'}\right) \Gamma,
    \label{EqCHI1}\\   
    0 & = -\sigma_w^2 \Gamma + \left(1 - a_p \right)\beta \sigma_\theta^2 - \frac{\overline{w'\theta'}}{\tau_R}.
    \label{EqCHIW1}
\end{align}
\end{subequations}
where $\overline{\varepsilon}$ and $\overline{\chi}_{\theta}$ denote the dissipation rates of TKE and $\sigma_\theta^2$, respectively. The variance of vertical velocity is $\sigma_w^2$. In Eq.~(\ref{EqCHIW1}), the parameter $a_p$ influences the buoyant contribution to the pressure-temperature interaction term; whereas, the last term of this equation is a parameterization of the turbulent-turbulent component of the pressure-temperature interaction. The return-to-isotropy time scale is denoted by $\tau_R$. Please refer to Appendix~1 for further technical details on the parameterization of pressure-temperature interaction. 

The Eqs.~(\ref{EqEDR1}), (\ref{EqCHI1}), and (\ref{EqCHIW1}) assume steady-state and horizontal homogeneity. Furthermore, the terms with secondary importance (e.g., turbulent transport) are neglected. Eqs.~(\ref{EqEDR1}) and (\ref{EqCHI1}) assume that production is locally balanced by dissipation. Please refer to Wyngaard \cite{wyngaard75} and Fitzjarrald \cite{fitzjarrald79} for further details. The celebrated `local scaling' hypothesis by Nieuwstadt \cite{nieuwstadt84} also utilizes these equations.  

\subsection{A Hybrid Length Scale}

In analogy to Prandtl's mixing length hypothesis (see \cite{arya01,monin71,weinstock81}), let us assume that $\sigma_w$ is a characteristic velocity scale for stably stratified flows. Further assume that $L_X$ and $L_X/\sigma_w$ are characteristic length and time scales, respectively. Then, the eddy diffusivity, the dissipation rates, and turbulent-turbulent component of the pressure-temperature interaction can be re-written as follows:    

\begin{subequations}
\begin{align}
    K_M & = c_1 \sigma_w L_X,
    \label{EqK3}\\
    \overline{\varepsilon} & = c_2 \frac{\sigma_w^2}{\left(\frac{L_X}{\sigma_w}\right)} = c_2 \frac{\sigma_w^3}{L_X},
    \label{EqEDR2}\\
    \overline{\chi}_{\theta} & = c_3 \frac{\sigma_\theta^2}{\left(\frac{L_X}{\sigma_w}\right)} = c_3 \frac{\sigma_w}{L_X} \sigma_\theta^2,
    \label{EqCHI2}\\
    \frac{\overline{w'\theta'}}{\tau_R} & = c_4 \frac{\overline{w'\theta'}}{\left(\frac{L_X}{\sigma_w}\right)} = -c_1c_4 \frac{\sigma_w^2}{Pr_t} \Gamma.
    \label{EqCHIW2}    
\end{align}
\end{subequations}
Here the unknown (non-dimensional) coefficients are denoted as $c_i$, where $i$ is an integer. The parameterizations for the dissipation rates (i.e., $\overline{\varepsilon}$ and $\overline{\chi}_{\theta}$) are further discussed in Section~\ref{Diss}. 

If we now make use of Eqs.~(\ref{EqK1}), (\ref{EqK2}), (\ref{EqPr}), (\ref{EqK3}), (\ref{EqEDR2}) and substitute all the terms of Eq.~(\ref{EqEDR1}), we arrive at: 
\begin{subequations}
\begin{align}
    c_2 \frac{\sigma_w^3}{L_X} &= c_1 \sigma_w L_X S^2 - c_1 \sigma_w L_X \left(\frac{\beta}{Pr_t}\right) \Gamma,\\
    \mbox{or, } c_2 \frac{\sigma_w^3}{L_X} &= c_1 \sigma_w L_X S^2 \left(1 - \frac{Ri_g}{Pr_t} \right).
    \label{EqL1}
\end{align}
\end{subequations}
By simplifying Eq.~(\ref{EqL1}), we get: 
\begin{subequations}
\begin{align}
    L_X &= \sqrt{\frac{c_2}{c_1}} \left(\frac{\sigma_w}{S} \right) \left( \frac{1}{\sqrt{1-Ri_g/Pr_t}} \right),
    \label{EqLH0}
    \\
    \mbox{or, } L_X &= c_H L_H \left( \frac{1}{\sqrt{1-Ri_g/Pr_t}} \right) = \frac{c_H L_H}{\sqrt{1 - R_f}}, 
    \label{EqLH1}
\end{align}
\end{subequations}
where $L_H (= \frac{\sigma_w}{S})$ is the Hunt length scale and $c_H$ is an unknown proportionality constant. The length scale equation, Eq.~(\ref{EqLH0}), was originally derived by Holtslag \cite{holtslag98}. 

The Hunt length scale is related to the so-called buoyancy length scale ($L_b$) as follows: 
\begin{equation}
    L_H = \left(\frac{\sigma_w}{S} \right) = \left(\frac{\sigma_w}{N} \right) \frac{N}{S} = \left(\frac{\sigma_w}{N} \right) \sqrt{Ri_g} = L_b \sqrt{Ri_g}.
    \label{LHLb}
\end{equation}
Thus, Eq.~(\ref{EqLH1}) can be re-written as:
\begin{align}
    L_X &= c_H L_b \left( \frac{\sqrt{Ri_g}}{\sqrt{1-Ri_g/Pr_t}} \right). 
\end{align}

If we substitute the individual terms of Eq.~(\ref{EqCHI1}) by utilizing Eqs.~(\ref{EqK2}), (\ref{EqPr}), (\ref{EqK3}), and (\ref{EqCHI2}), we get:
\begin{align}
    c_3 \frac{\sigma_w}{L_X} \sigma_\theta^2 = 2 c_1 \frac{\sigma_w L_X}{Pr_t} \Gamma^2.
    \label{EqL3}
\end{align}
Simplification of this equation leads to:
\begin{subequations}
\begin{align}
    L_X &= \sqrt{\frac{c_3}{2c_1}} \left(\frac{\sigma_\theta}{\Gamma}\right) \sqrt{Pr_t},
    \\
    \mbox{or, } L_X &= c_E L_E \sqrt{Pr_t}. 
    \label{EqLE1}
\end{align}
\end{subequations}
where $L_E \left(=\sigma_\theta/\Gamma \right)$ is the Ellison length scale and $c_E$ is an unknown (nondimensional) coefficient. 

We would like to point out that in the appendices of Basu et al.~\cite{basu21b,basu21a} we have summarized the characteristics of Hunt, buoyancy, Ellison, Bolgiano, Ozmidov, and several other length scales. For brevity, we do not repeat them here.  

\subsection{Ratios of Length Scales}

By comparing Eq.~(\ref{EqLH1}) with Eq.~(\ref{EqLE1}), it is rather straightforward to derive: 
\begin{subequations}
\begin{align}
    Pr_t  &=  \left(\frac{c_H L_H}{c_E L_E}\right)^2 + Ri_g,\\
    \mbox{or, } \frac{L_H^2}{L_E^2} &= \frac{\left(Pr_t - Ri_g \right)}{c_P},
    \label{Pr2}
\end{align}
\end{subequations}
where $c_P = \frac{c_H^2}{c_E^2}$. Using Eq.~(\ref{LHLb}), this equation can be re-written as follows: 
\begin{equation}
    \frac{L_b^2}{L_E^2} = \frac{\left(Pr_t - Ri_g \right)}{c_P Ri_g} = \frac{\left(1-R_f\right)}{c_P R_f}.
    \label{LbLE1}
\end{equation}

An alternative expression for $\left(\frac{L_b^2}{L_E^2}\right)$ can be found if we use Eqs.~(\ref{EqK2}), (\ref{EqPr}), (\ref{EqK3}), and (\ref{EqCHIW2}) to substitute the individual terms of Eq.~(\ref{EqCHIW1}) as follows:
\begin{subequations}
\begin{align}
    -c_1 c_4 \frac{\sigma_w^2}{Pr_t} \Gamma &=  -\sigma_w^2 \Gamma + \left(1 - a_p \right)\beta \sigma_\theta^2,\\
    \mbox{or, } \left(1 - \frac{c_5}{Pr_t}\right) \sigma_w^2 \Gamma &=  \left(1 - a_p \right)\beta \sigma_\theta^2,\\
    \mbox{or, } \left(1 - \frac{c_5}{Pr_t}\right) L_b^2 &= \left(1 - a_p \right) L_E^2,\\
    \mbox{or, } \frac{L_b^2}{L_E^2} &= \frac{\left(1 - a_p \right)}{\left(1 - \frac{c_5}{Pr_t}\right)},
    \label{LbLE2}
\end{align}
\end{subequations}
where $c_5 (= c_1 c_4)$ is an unknown proportionality constant. 

\subsection{Derivation of Prandtl Number}

By equating Eq.~(\ref{LbLE1}) and Eq.~(\ref{LbLE2}), we immediately get the following quadratic equation: 
\begin{equation}
    Pr_t^2 - \left[c_5 + Ri_g + \left(1 - a_p\right) c_P Ri_g \right] Pr_t + c_5 Ri_g = 0.
    \label{QuadPr}
\end{equation}
Since $Pr_t = Pr_{t0}$ for neutral conditions ($Ri_g = 0$), via Eq.~(\ref{QuadPr}), we find: 
\begin{equation}
c_5 = Pr_{t0}.
\label{c5}
\end{equation}
The roots of Eq.~(\ref{QuadPr}) are: 
\begin{equation}
    Pr_t = \frac{X \pm \sqrt{X^2 - 4 Pr_{t0} Ri_g}}{2},
    \label{PrTFinal}
\end{equation}
where, $X = \left[Pr_{t0} + Ri_g + \left(1 - a_p\right) c_P Ri_g \right]$. Only the larger root is physically meaningful. Eq.~(\ref{PrTFinal}) includes three unknown parameters (i.e., $Pr_{t0}$, $a_p$, and $c_P$). Similarity theory can be used to estimate $c_P$ (discussed in the following section). However, $Pr_{t0}$ and $a_p$ must be prescribed.  

We would like to emphasize that Eq.~(\ref{PrTFinal}) is a closed form analytical solution for the stability-dependence of $Pr_t$. It is derived directly from the budget equations without any additional simplification. Since our derivation makes use of certain length scale ratios (LSRs), we refer to our proposed approach as the LSR formulation.

\section{Estimation of Unknown Coefficients}

For near-neutral conditions, Eqs.~(\ref{EqLH1}) and (\ref{EqLE1}) simplify to the following expressions, respectively:
\begin{subequations}
\begin{align}
    L_X & \approx c_H \frac{\sigma_w}{S},\\ 
    L_X & \approx c_E \frac{\sigma_\theta}{\Gamma}\sqrt{Pr_{t0}}. 
\end{align}
In order to be consistent with the logarithmic velocity profile in the surface layer, $L_X$ should be equal to $\kappa z$ in the surface layer, where $\kappa$ is the von K\'{a}rm\'{a}n constant. Therefore, 
\begin{align}
    c_H & \approx \frac{\kappa z S}{\sigma_w},\\
    c_E & \approx \frac{\kappa z \Gamma}{\sqrt{Pr_{t0}}\sigma_\theta}.
\end{align}
Numerous studies reported that $\sigma_w = c_w u_*$ and $\sigma_\theta = c_\theta \theta_*$ in near-neutral stratified surface layer. The surface friction velocity and temperature scale are denoted by $u_*$ and $\theta_*$, respectively. Thus, we get:
\begin{align}
    c_H & \approx \frac{\kappa z S}{c_w u_*} = \frac{1}{c_w}, \label{cH}\\
    c_E & \approx \frac{\kappa z \Gamma}{\sqrt{Pr_{t0}} c_{\theta} \theta_*} = \frac{\sqrt{Pr_{t0}}}{c_\theta}. \label{cE}
\end{align}
Please note that the non-dimensional velocity gradient, $\left(\kappa z S/u_*\right)$, equals to unity according to the logarithmic law of the wall. Whereas, the non-dimensional temperature gradient, $\left(\kappa z \Gamma/\theta_*\right)$, equals to $Pr_{t0}$.  
\end{subequations}

By using Eqs.~(\ref{EqK1}), (\ref{EqK3}), and (\ref{EqLH1}), we can expand the along-wind momentum flux as follows:
\begin{subequations}
\begin{align}
\overline{u'w'} & = - c_1 c_H \sigma_w^2 \frac{1}{\sqrt{1-Ri_g/Pr_t}},
\label{Ruw1}
\end{align}
Thus, the normalized momentum flux can be written as:
\begin{align}
    R_{uw} & = \left(\frac{\overline{u'w'}}{\sigma_w^2}\right) = -\frac{c_1 c_H}{\sqrt{1 - Ri_g/Pr_t}}.
\label{Ruw2}
\end{align}
For neutral condition, $R_{uw}$ simplifies to: $R_{uw0} = -c_1 c_H$. Since, $\sigma_w = c_w u_*$, we get: 
\begin{align}
    R_{uw0} = -\frac{1}{c_w^2} = -c_1 c_H. 
\label{Ruw3}
\end{align}
Since, $c_H \approx \frac{1}{c_w}$, the unknown coefficient $c_1$ is also approximately equal to $\frac{1}{c_w}$. 
\end{subequations}
Typical values of $R_{uw0}$ are documented in Table~\ref{T1}.

From Eqs~(\ref{EqLH1}), (\ref{EqLE1}), (\ref{Pr2}), (\ref{LbLE2}), (\ref{c5}), (\ref{cH}), and (\ref{cE}), via simple algebraic calculations, we can write all the unknown $c_i$ coefficients as functions of $c_w$, $c_\theta$, and $Pr_{t0}$ as follows: 
\begin{subequations}
\begin{equation}
c_1 = c_H = \frac{1}{c_w},    
\end{equation}
\begin{equation}
c_2 = c_H^3 = \frac{1}{c_w^3},    
\end{equation}
\begin{equation}
c_3 = \frac{2 Pr_{t0}}{c_w c_\theta^2},    
\end{equation}
\begin{equation}
c_4 = Pr_{t0} c_w,    
\end{equation}
\begin{equation}
\mbox{and recall that}\hspace{0.1in} c_5 = Pr_{t0}.    
\end{equation}
\label{Eqcoeffs}
\end{subequations}

In the literature, the most commonly reported values of $c_w$ range from 1.25--1.30 \cite{arya01,kaimal94,nieuwstadt84,sorbjan89}. Similarly, $c_\theta$ values vary approximately from 1.8 to 2.0 \cite{kaimal94,sorbjan89}. In a few publications, somewhat different values were also reported (e.g., \cite{lumley64,wilson08}). In Table~\ref{T1}, we have computed $c_i$ and other coefficients for a few combinations of $Pr_{t0}$, $c_w$, and $c_\theta$. 

\begin{table*}[t]
\caption{Statistics associated the proposed LSR Model}
\label{T1}       
\begin{tabular}{lll|llllllllll}
\hline\noalign{\smallskip}
\multicolumn{3}{c|}{Prescribed} & \multicolumn{10}{c}{Estimated} \\
\hline\noalign{\smallskip}
$Pr_{t0}$ & $c_w$ & $c_\theta$ & $c_H$ & $c_E$ & $c_P$ & $c_1$ & $c_2$ & $c_3$ & $c_4$ & $c_5$ & $R_{uw0}$ & $R_{w\theta0}$\\
\noalign{\smallskip}\hline\noalign{\smallskip}
0.74 & 1.25 & 1.80 & 0.80 & 0.48 & 2.80 & 0.80 & 0.51 & 0.37 & 0.93 & 0.74 & -0.64 & -0.44\\
0.74 & 1.30 & 2.00 & 0.77 & 0.43 & 3.20 & 0.77 & 0.46 & 0.28 & 0.96 & 0.74 & -0.59 & -0.38\\
0.85 & 1.25 & 1.80 & 0.80 & 0.51 & 2.44 & 0.80 & 0.51 & 0.42 & 1.06 & 0.85 & -0.64 & -0.44\\
0.85 & 1.30 & 2.00 & 0.77 & 0.46 & 2.78 & 0.77 & 0.46 & 0.33 & 1.11 & 0.85 & -0.59 & -0.38\\
0.85 & 1.05 & 2.00 & 0.95 & 0.46 & 4.27 & 0.95 & 0.86 & 0.40 & 0.89 & 0.85 & -0.91 & -0.48\\
\noalign{\smallskip}\hline
\end{tabular}
\end{table*}

\section{Parameterizations of Dissipation Rates}
\label{Diss}

\subsection{Energy Dissipation Rate}

The energy dissipation rate is commonly parameterized as follows \cite{mellor82}: 
\begin{equation}
    \overline{\varepsilon} = \frac{q^3}{B_1 L_M},
    \label{EqEDRTKE}
\end{equation}
where $q^2$ is twice TKE. $L_M$ is known as the master length scale and $B_1$ is a constant coefficient. In this study, following Townsend~\cite{townsend58}, we use Eq.~(\ref{EqEDR2}) as an alternative parameterization for $\overline{\varepsilon}$  which makes use of $\sigma_w^3$ instead of $q^3$. Using Eqs.~(\ref{EqLH1}), and (\ref{Eqcoeffs}), we can re-write this parameterization as follows: 
\begin{subequations}
\begin{widetext}
\begin{equation}
\overline{\varepsilon} = \left(\frac{c_2}{c_H} \right)\sigma_w^2 S \sqrt{1 - Ri_g/Pr_t} = \left(\frac{1}{c_w^2} \right)\sigma_w^2 S \sqrt{1 - Ri_g/Pr_t}. 
\end{equation}
\end{widetext}
If the value of $c_w$ is approximately in the range of 1.25--1.30 (refer to Table~{T1}), for small values of $Ri_g$ (i.e., weakly stable conditions), we get:
\begin{equation}
\overline{\varepsilon} = 0.60 \sigma_w^2 S.     
\label{EqEDRHunt}
\end{equation}
It is important to note that Eq.~(\ref{EqEDRHunt}) (with an unknown proportionality constant) was originally proposed by Hunt~\cite{hunt88} using heuristic arguments. He hypothesized that the energy dissipation in weakly/moderately stably stratified flows is dictated by mean shear ($S$) and root-mean-square value of vertical velocity fluctuations (i.e., $\sigma_w$) which is the characteristic velocity scale in the direction of $S$. Later on Schumann and Gerz~\cite{schumann95} analyzed various observational and simulation datasets and validated Hunt's parameterization (see their Figure~1). More recently, Basu et al.~\cite{basu21a} utilized a database of direct numerical simulations and found: 
\begin{equation}
\overline{\varepsilon} = 0.23 \overline{e} S = 0.63 \sigma_w^2 S, 
\label{EqEDRBasu21a}
\end{equation}
for $0 < Ri_g < 0.2$. TKE is denoted by $\overline{e}$. 
\end{subequations}
It is remarkable that the DNS-based empirical formulation of \cite{basu21a} is virtually identical to our analytical prediction, i.e., Eq.~(\ref{EqEDRHunt}). However, we are unable to ascertain the validity of either Eq.~(\ref{EqEDRHunt}) or Eq.~(\ref{EqEDRBasu21a}) for $Ri_g > 0.2$. We will discuss more on this issue in Section~\ref{disc}.  

The exact value of $B_1$ in Eq.~(\ref{EqEDRTKE}) is not settled in the literature. Over the years, a number of researchers estimated its value from diverse observational and simulated datasets; see a brief summary in Table~\ref{B1}. By combining the analytical results from the present study with the DNS results from Basu et al.~\cite{basu21a}, we can also estimate $B_1$ as follows. 
From Eq.~(\ref{EqEDRBasu21a}), for $0 < Ri_g < 0.2$, we can write:
\begin{equation}
    \overline{e} = \frac{q^2}{2} = \left(\frac{0.63}{0.23}\right)\sigma_w^2 = 2.74 \sigma_w^2. 
\end{equation}
Next, if we assume our proposed length scale ($L_X$) is equal to the master length scale ($L_M$), then from Eqs.~(\ref{EqLH1}) and (\ref{EqEDRTKE}), we get: 
\begin{equation}
    B_1 = \frac{q^3}{\overline{\varepsilon} L_X} = \frac{ (2\times 2.74)^{3/2} \sigma_w^3}{\left(0.63 \sigma_w^2 S\right) \left(c_H \sigma_w/S\right)} = 25.5.   
\end{equation}
Here we have assumed $c_H = 0.8$ and $\sqrt{1-Ri_g/Pr_t} \approx 1$ for small values of $Ri_g$. Clearly, our estimated value of $B_1$ agrees reasonably well with some of the published studies; however, it is significantly higher than the widely used value of 16.6. Please note that due to a missing multiplying coefficient of value 2.1, Basu et al.~\cite{basu21a} incorrectly reported $B_1$ = 12.3 instead of 25.8.  

\begin{table}[t]
\caption{Published values of $B_1$ coefficient}
\label{B1}       
\begin{tabular}{ll}
\noalign{\smallskip}\hline
Study & $B_1$\\
\noalign{\smallskip}\hline
Mellor and Yamada \cite{mellor82} & 16.6\\
Enger \cite{enger86} & 27.0\\
Andr\'{e}n and Moeng \cite{andren93} & 27.4\\
Nakanishi \cite{nakanishi01} & 24.0\\
Janji\'{c} \cite{janjic02} & 11.9\\
Cheng et al. \cite{cheng02} & 19.3\\
Basu et al. \cite{basu21a} & 25.8\\
\hline\noalign{\smallskip}
\end{tabular}
\end{table}

\subsection{Dissipation Rate of Temperature Variance}

Once again, following Townsend~\cite{townsend58}, we parameterized the dissipation rate of temperature variance ($\overline{\chi}_\theta$) by Eq.~(\ref{EqCHI2}). Combining this equation with Eq.~(\ref{EqLH1}), Eq.~(\ref{EqL3}), and Eqs.~(\ref{Eqcoeffs}), we get: 
\begin{widetext}
\begin{equation}
    \overline{\chi}_\theta = \left( \frac{2 c_1 c_H}{Pr_t}\right)  \frac{\left(\frac{\sigma_w^2}{S}\right)\Gamma^2}{\sqrt{1 - Ri_g/Pr_t}} = \left( \frac{2 c_H^2}{Pr_t}\right)  \frac{\left(\frac{\sigma_w^2}{S}\right)\Gamma^2}{\sqrt{1 - Ri_g/Pr_t}}.
\end{equation}
\end{widetext}
For small values of $Ri_g$, we can assume $Pr_t \approx 0.85$. As before, if we also consider $c_H = 0.8$, we arrive at: $\overline{\chi}_\theta \approx 1.51\left(\frac{\sigma_w^2}{S}\right)\Gamma^2$. Almost the same formulation was reported by Basu et al.~\cite{basu21b} based on their analysis of a DNS database. For $0 < Ri_g < 0.2$, they found: $\overline{\chi}_\theta=1.47\left(\frac{\sigma_w^2}{S}\right)\Gamma^2$. 

In summary of this section, we can state that our analytical formulations of dissipation rates are very reliable for $0 < Ri_g < 0.2$. However, more research will be needed for their rigorous validation for the very stable regime (i.e., $Ri_g > 0.2$). 

\section{Results}

\subsection{Turbulent Prandtl Number}

Our proposed formulation for the turbulent Prandtl number, Eq.~(\ref{PrTFinal}), contains 3 unknown coefficients: $Pr_{t0}$, $a_p$, and $c_P$. Based on the discussion in the Introduction, in this study, we have opted to use $Pr_{t0}$ = 0.85. The value of $c_P$ is selected from Table~\ref{T1}; it is evident that it should vary within a range of 2.4--4.3 for typical values of $c_w$ and $c_\theta$. The parameter $a_p$ is discussed in Appendix~1. 

In Fig.~\ref{FigPr}, the predictions from our LSR approach are reported for various combinations of $a_p$ and $c_P$. In addition to $Pr_t$, we have also reported the stability-dependence of $R_f$. The results are sensitive to $a_p$ values for $Ri_g > 0.1$. It is encouraging to see that the predictions are qualitatively in agreement with the published observations. They are also in-line with the predictions from the co-spectral budget (CSB; \cite{katul14}) and energy- and flux-budget (EFB; \cite{zilitinkevich13}) approaches.

We would like to emphasize out that Eq.~(\ref{PrTFinal}) and Eq.~(\ref{CSB-PrTFinal}) in Appendix~2 have nearly identical mathematical form despite the fundamental differences in the LSR and CSB approaches. The CSB approach includes prescribed coefficients from Kolmogorov-Obukhov-Corrsin hypotheses and from a parameterization of the pressure-temperature decorrelation (refer to Appendix~2); they are all lumped into a variable called $\omega^{CSB}$ in Eq.~(\ref{CSB-PrTFinal}). However, it does not consider the buoyancy-turbulence interaction term in the sensible heat flux equation. Thus, Eq.~(\ref{CSB-PrTFinal}) does not include the $a_p$ parameter. In contrast, the LSR approach largely depends on $c_w$ and $c_\theta$ coefficients (combined into the $c_P$ coefficient) in addition to $a_p$. These coefficients are integral part of surface layer similarity theory for near-neutral conditions. Furthermore, by construction, the CSB approach assumes $Pr_{t0} = 1$. Whereas, in the case of the LSR approach, $Pr_{t0}$ is assumed to be equal to 0.85. 

For very stable condition (i.e., $Ri_g \gg 1$), Eq.~(\ref{PrTFinal}) is simplified to: 
\begin{equation}
    Pr_t \approx \left(1 + \left(1 - a_p\right) c_P \right) Ri_g = \frac{Ri_g}{R_{f\infty}}.
    \label{EqPrInfty}
\end{equation}
In contrast, Eq.~(\ref{CSB-PrTFinal}) from the CSB approach leads to: 
\begin{equation}
    Pr_t \approx \omega^{CSB} Ri_g \approx 4 Ri_g.
    \label{CSB-EqPrInfty}
\end{equation}
Thus, the CSB approach predicts $R_{f\infty} \approx 0.25$. On the other hand, for $a_p$ = 0 and $c_P$ = 4.27, $R_{f\infty}$ equals to 0.19 for the LSR approach. However, for $a_p$ = 0.5 and $c_P$ = 2.4, $R_{f\infty}$ increases to 0.46. In the literature (see \cite{ellison57}, \cite{grachev13}, \cite{townsend58}, \cite{yamada75}), $R_{f\infty}$ has been reported to be within the limits of 0.15 and 0.5; both the LSR-based and CSB-based predictions are in this range.

\begin{figure*}[ht]
  \centerline{
  \includegraphics[height=2.2in]{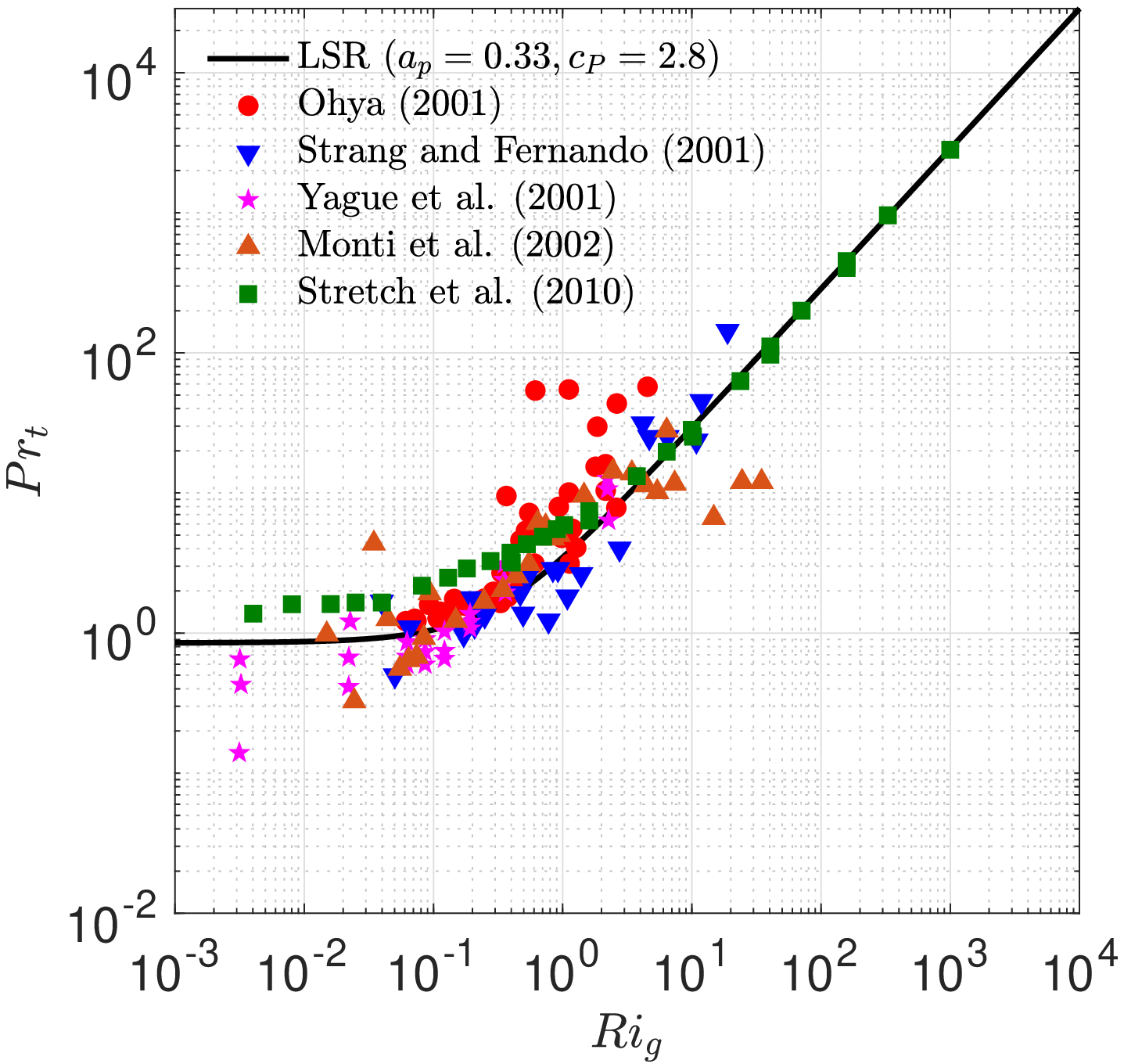}
  \hspace{0.3in}
  \includegraphics[height=2.2in]{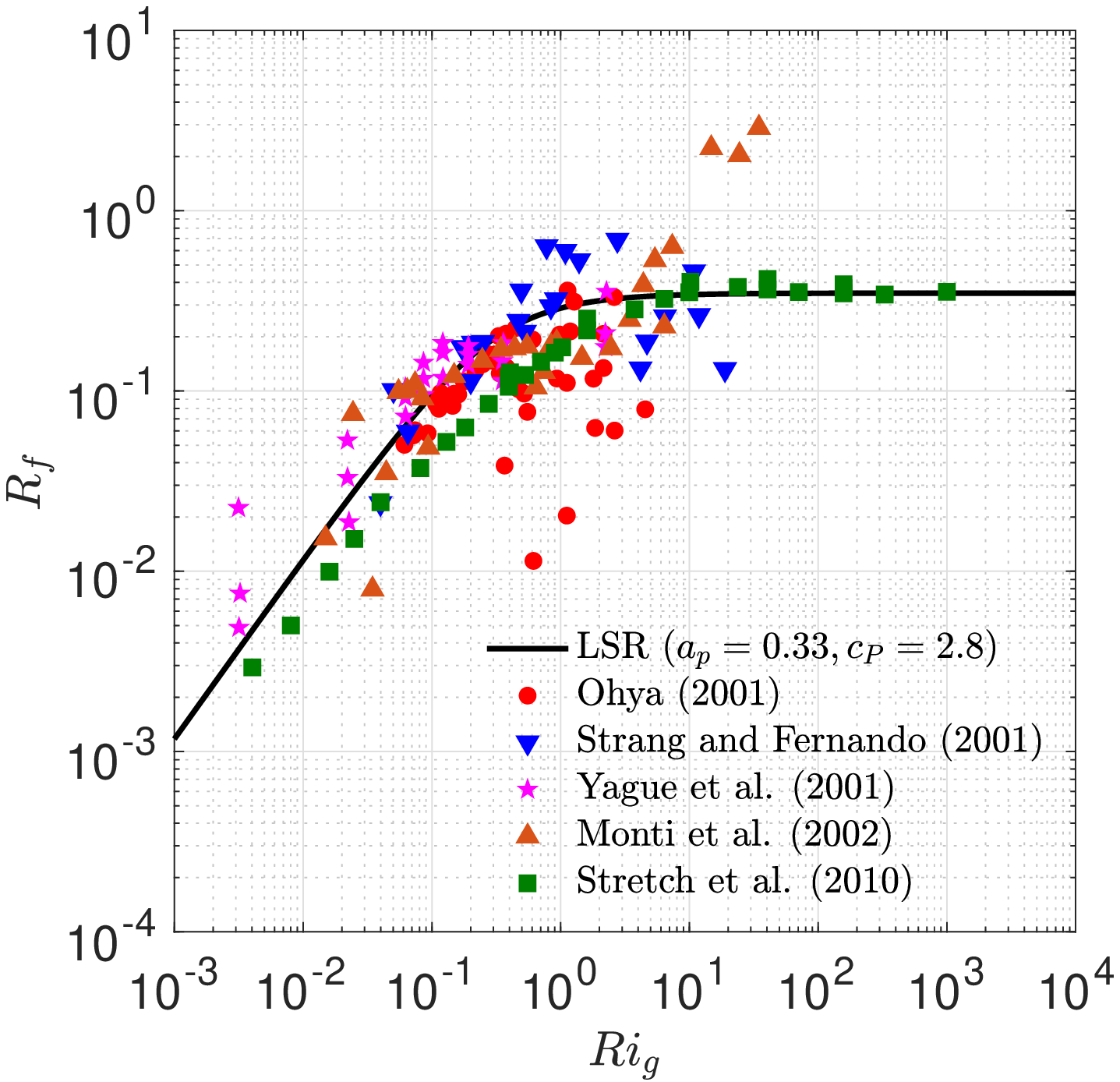}}
  \centerline{
  \includegraphics[height=2.2in]{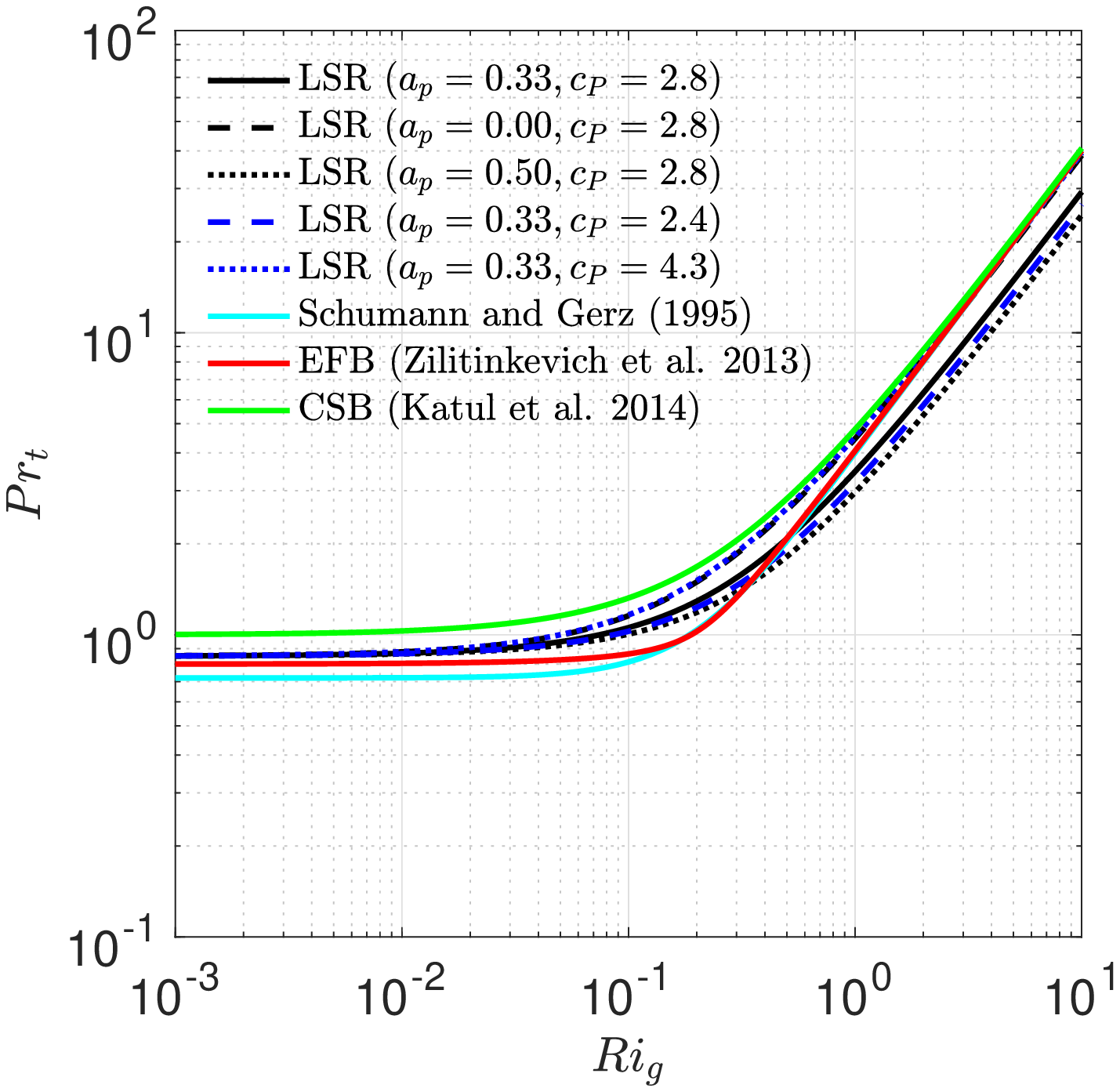}
  \hspace{0.3in}
  \includegraphics[height=2.2in]{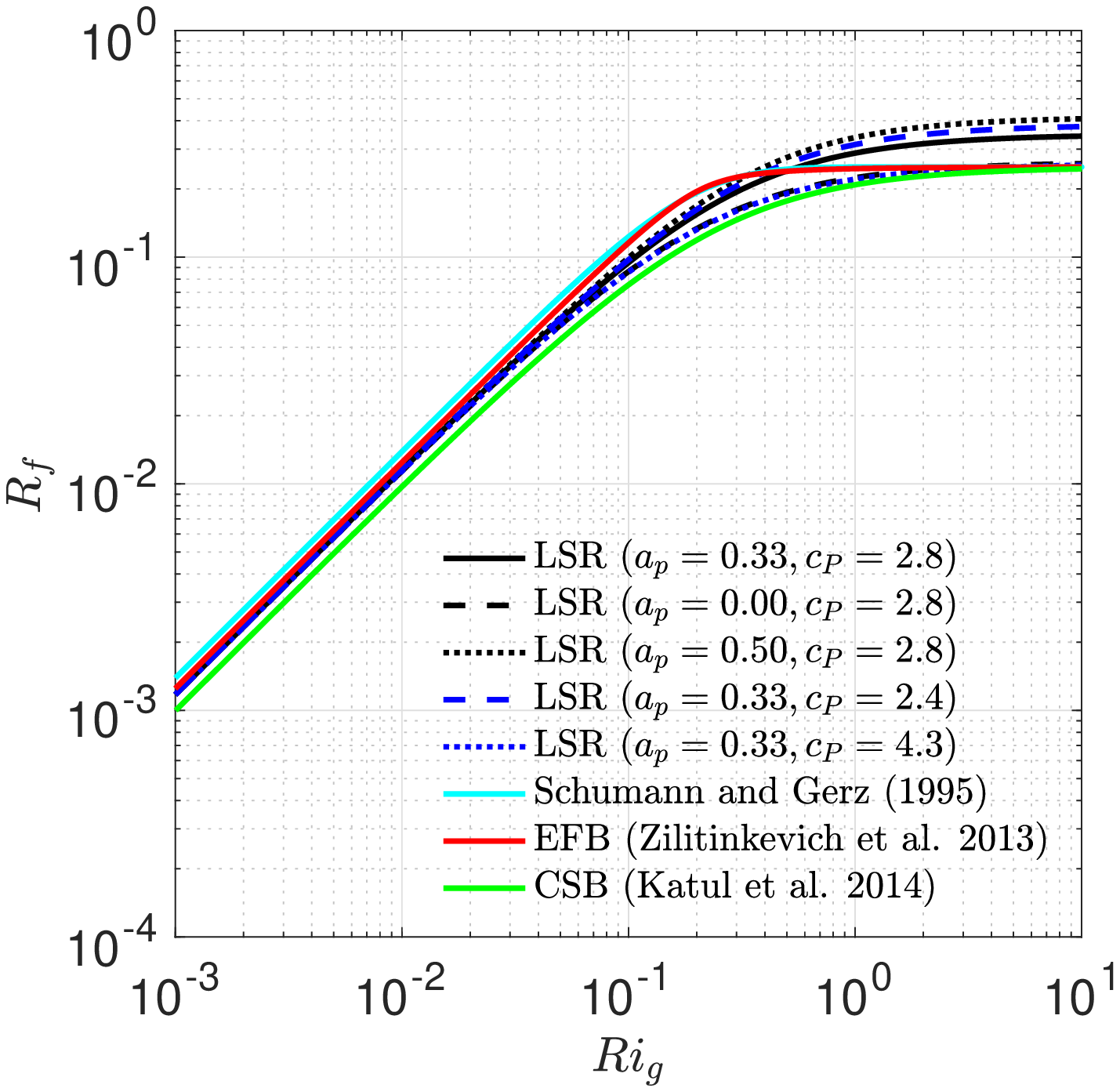}}
  \caption{The dependence of $Pr_t$ (left panel) and $R_f$ (right panel) on $Ri_g$. As a default, the length scale ratio (LSR) approach assumes $Pr_{t0} = 0.85$, $a_p$ = 0.33, and $c_P$ = 2.8. In the top panels, published data from various sources \cite{monti02,ohya01,strang01,stretch10,yague01} are overlaid. The sensitivities of the LSR-based predictions with respect to $a_p$ and $c_P$ coefficients are documented in the bottom panels. The predictions from Schumann and Gerz~\cite{schumann95}, Zilitinkevich et al.~\cite{zilitinkevich13}, and Katul et al.~\cite{katul14} are also shown in these panels for comparison.}
\label{FigPr}
\end{figure*}

\subsection{Normalized Variances and Fluxes}

In the literature, there is no consensus regarding the exact stability-dependence of a few normalized variables. Different formulations (e.g., \cite{li16}, \cite{zilitinkevich13}) predict different trends. The LSR approach allows us to independently predict some of these ratios without further approximations as elaborated below.    

\subsubsection{Ratio of Turbulent Potential and Kinetic Energies}

\begin{figure*}[ht!]
 \centerline{
  \includegraphics[height=1.75in]{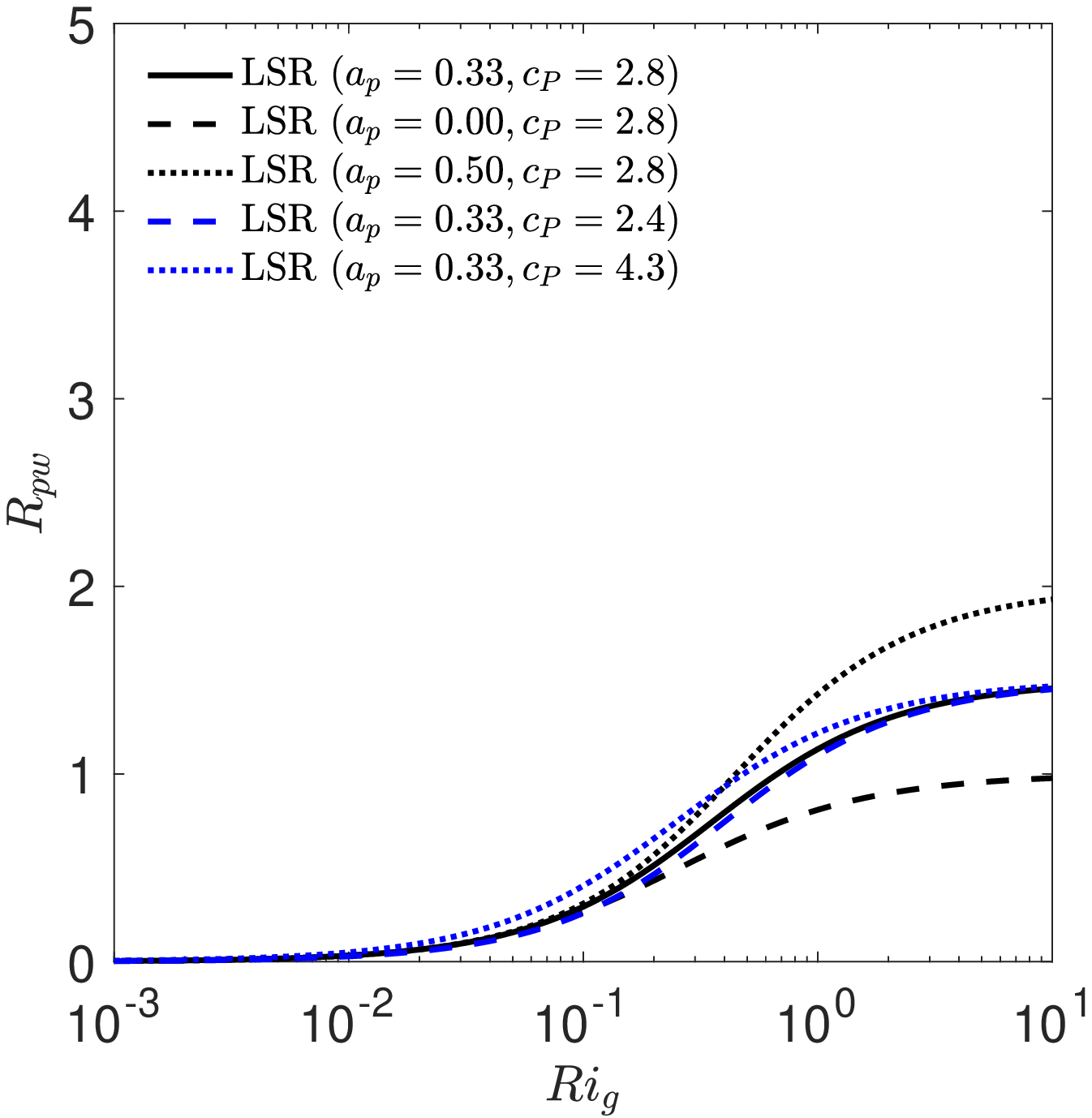}
  \hspace{0.3in}
  \includegraphics[height=1.75in]{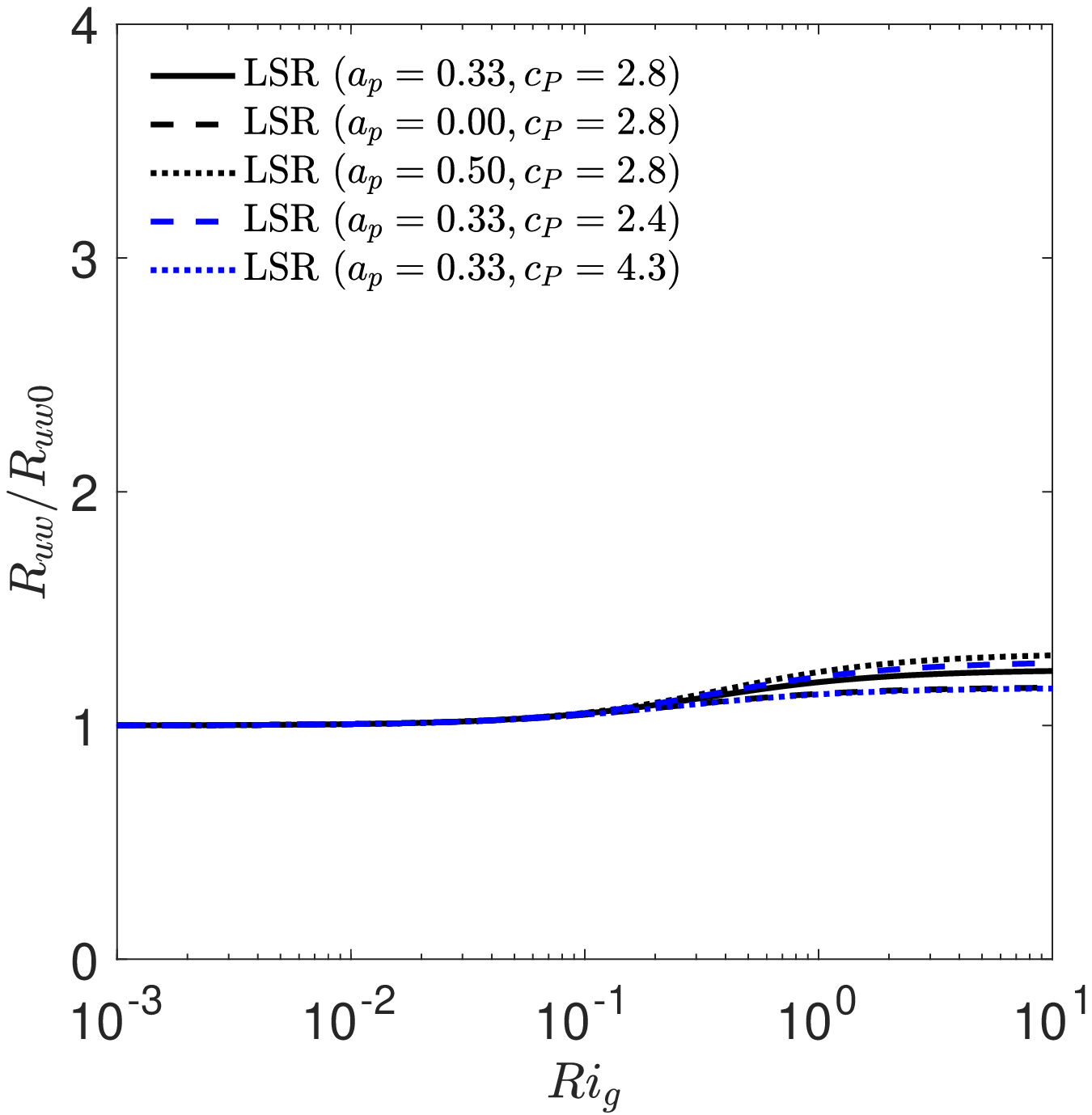}
  \hspace{0.3in}
  \includegraphics[height=1.75in]{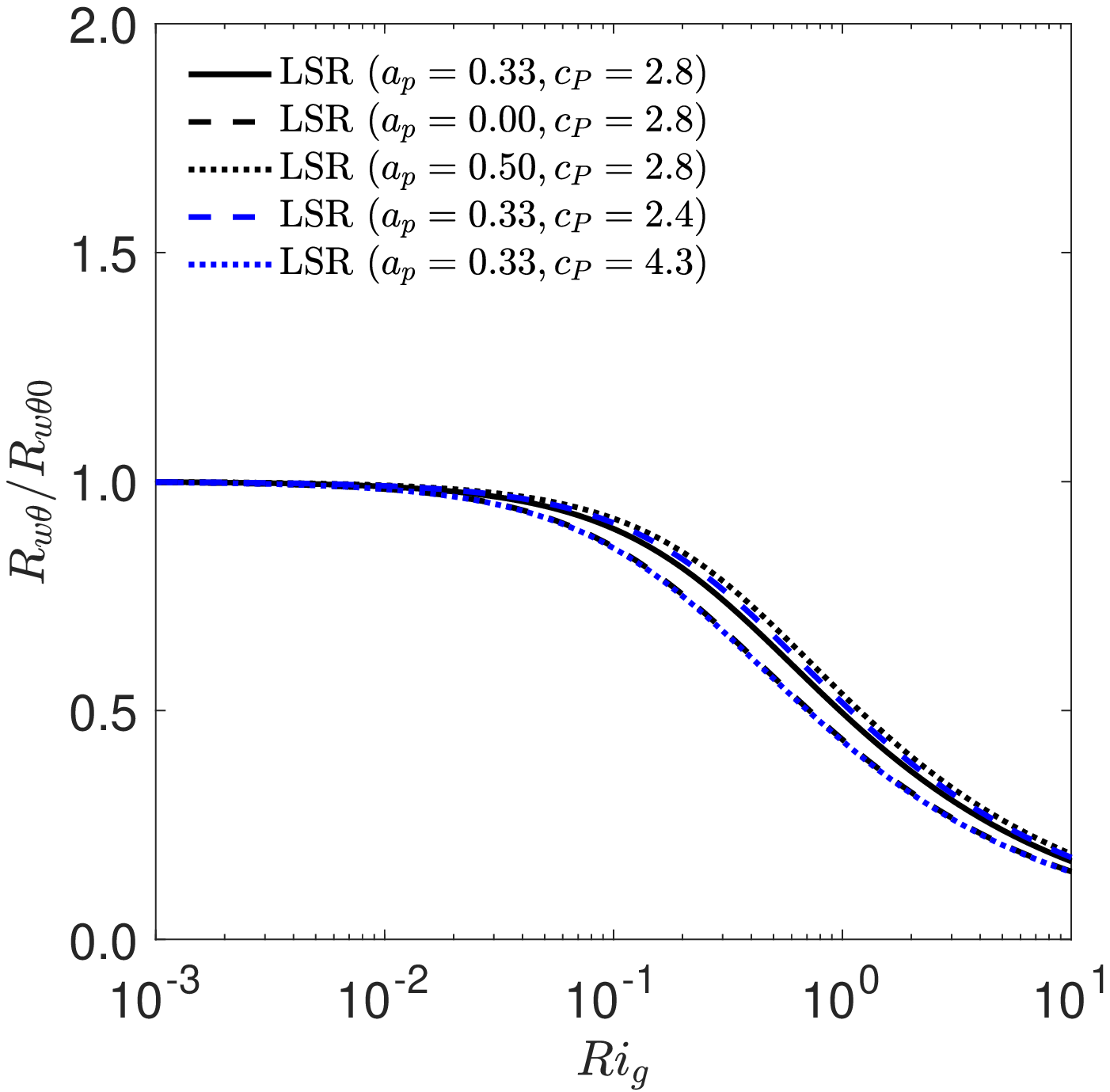}}
  \caption{The dependence of $R_{pw}$ (top panel), normalized $R_{uw}$ (middle panel), and normalized $R_{w\theta}$ (bottom panel), on $Ri_g$. As a default, the length scale ratio (LSR) approach assumes $Pr_{t0} = 0.85$, $a_p$ = 0.33, and $c_P$ = 2.8. The sensitivities of the LSR-based predictions with respect to $a_p$ and $c_P$ coefficients are documented in all the panels.}
\label{FigRatio}
\end{figure*}

We first consider the ratio of the turbulent potential energy (TPE; denoted as $\overline{e}_p$) and the vertical component of TKE (i.e., $\overline{e}_w$). These variables are commonly written as \cite{li16}: 
\begin{subequations}
\begin{align}
    \overline{e}_p & = \left(\frac{\beta}{N} \right)^2 e_T,
    \label{ep}
    \\
    \overline{e}_w & = \frac{\sigma_w^2}{2},
    \label{ew}
\end{align}
\end{subequations}
where $\overline{e}_T = \frac{\sigma_\theta^2}{2}$. By using the definition of the Ellison length scale ($L_E$), we can re-write $\overline{e}_p$ as follows: 
\begin{equation}
    \overline{e}_p = \frac{1}{2} N^2 L_E^2.
\end{equation}
Thus, the ratio of $\overline{e}_p$ and $\overline{e}_w$ is simply: 
\begin{subequations}
\begin{equation}
    R_{pw} = \frac{\overline{e}_p}{\overline{e}_w} = \frac{N^2 L_E^2}{\sigma_w^2} = \frac{L_E^2}{L_b^2}. 
\end{equation}
By making use of Eq.~(\ref{LbLE1}), we can re-write $R_{pw}$ as follows:
\begin{equation}
    R_{pw} = \frac{c_P Ri_g}{\left(Pr_t - Ri_g\right)} = \frac{c_P R_f}{\left(1-R_f\right)}. 
    \label{LSR-Rpw1}
\end{equation}
\end{subequations}
In the top panel of Fig.~\ref{FigRatio}, the dependence of $R_{pw}$ on $Ri_g$ is shown. Clearly, $R_{pw}$ is strongly influenced by $a_p$ for $Ri_g > 0.2$. In contrast, somewhat surprisingly, $R_{pw}$ is not very sensitive to the coefficient $c_P$. In the denominator of $R_{pw}$, the term $(Pr_t - Rig)$ appears which strongly depends on $c_P$. It effectively cancels out the influence of $c_P$ in the numerator of $R_{pw}$. 

\subsubsection{Normalized Momentum Flux}

The formulations for $R_{uw}$ and $R_{uw0}$ are derived earlier in Eqs.~(\ref{Ruw2}) and (\ref{Ruw3}), respectively. 
Hence, their ratio becomes:
\begin{align}
    \frac{R_{uw}}{R_{uw0}} & = \frac{1}{\sqrt{1 - Ri_g/Pr_t}} = \frac{1}{\sqrt{1-R_f}}.
    \label{LSR-Ruw}
\end{align}
The dependence of the normalized momentum flux on $Ri_g$ is shown in the middle panel of Fig.~\ref{FigRatio}. It is marginally sensitive to $a_p$ and $c_P$.  

\subsubsection{Normalized Correlation of $w$ and $\theta$}

Similar to the momentum flux expression, the sensible heat flux can be re-written using Eqs.~(\ref{EqK2}), (\ref{EqPr}), (\ref{EqK3}), and (\ref{EqLE1}) as follows: 
\begin{subequations}
\begin{align}
    \overline{w'\theta'} & = - c_1 c_E \sigma_w \sigma_\theta \frac{1}{\sqrt{Pr_t}}.
\end{align}
Hence, the correlation between $w$ and $\theta$ becomes:
\begin{align}
    R_{w\theta} & = \left(\frac{\overline{w'\theta'}}{\sigma_w \sigma_\theta}\right) = - \frac{c_1 c_E}{\sqrt{Pr_t}}.
\end{align}
For neutral condition, we have $R_{w\theta 0} = -\frac{c_1 c_E}{\sqrt{Pr_{t0}}}$. So, the normalized correlation can be written as: 
\begin{align}
    \frac{R_{w\theta}}{R_{w\theta 0}} & = \sqrt{\frac{Pr_{t0}}{Pr_t}}.
    \label{LSR-RwT}
\end{align}
\end{subequations}
Typical values of $R_{w\theta 0}$ are documented in Table~\ref{T1}. The normalized correlations are plotted in the right panel of Fig.~\ref{FigRatio}. Similar to the normalized momentum flux, this ratio is also very weakly dependent on $a_p$ and $c_P$.  

\subsubsection{Comparison of Different Theoretical Approaches}
\label{Comparison}

\begin{figure*}[ht!]
 \centerline{
  \includegraphics[height=1.5in]{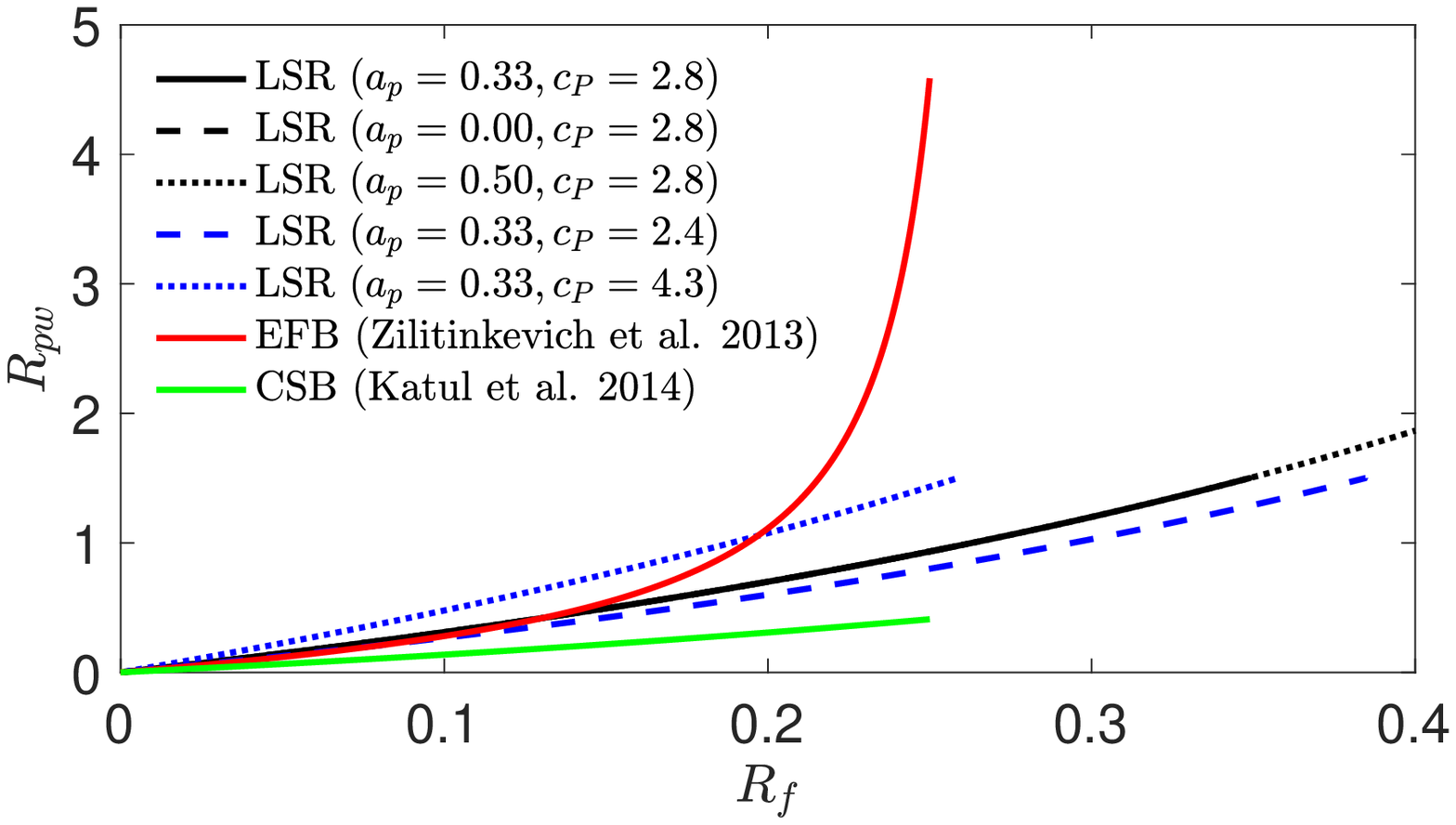}
  \hspace{0.3in}
  \includegraphics[height=1.5in]{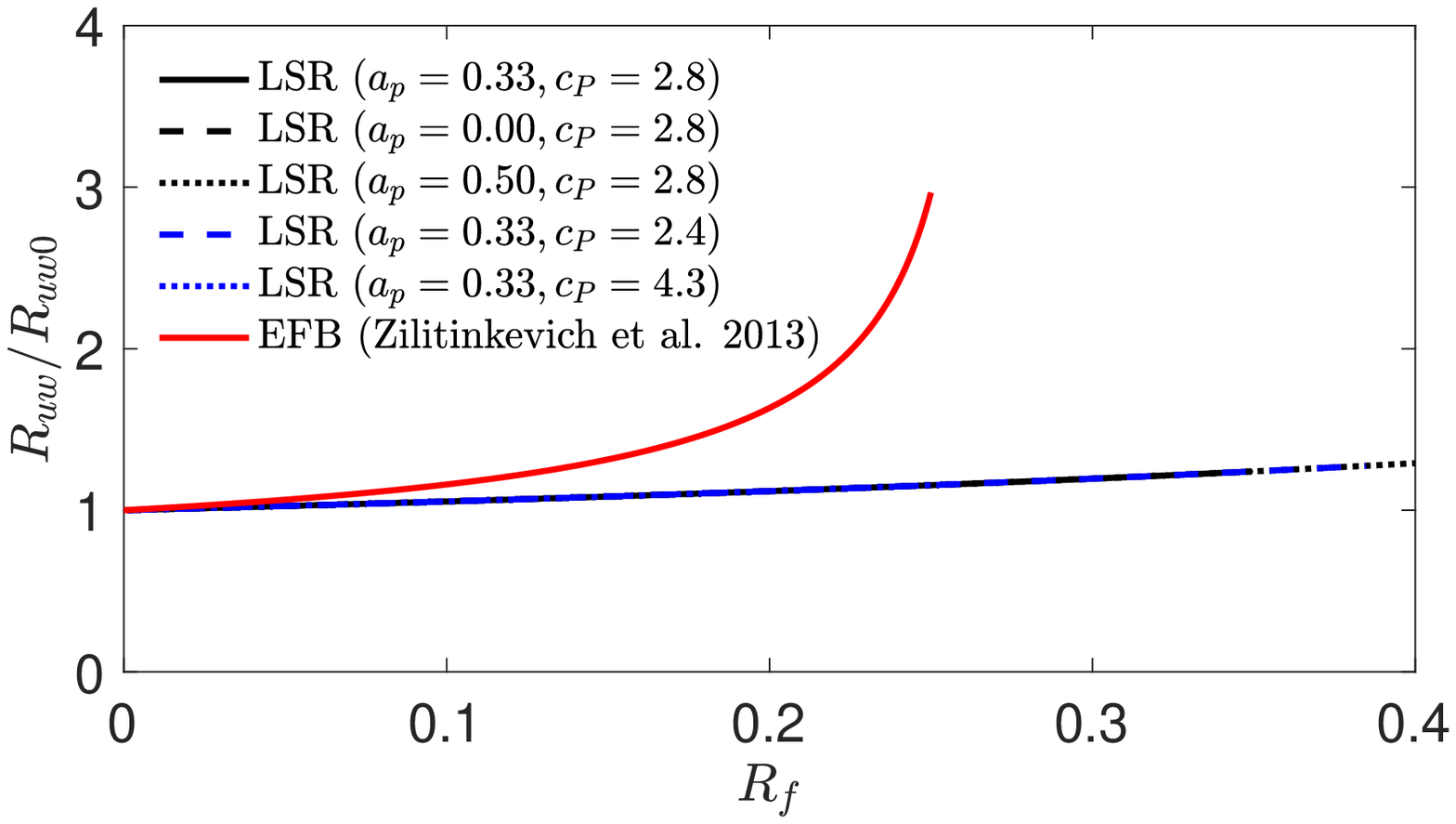}}
  \caption{The dependence of $R_{pw}$ (left panel) and normalized $R_{uw}$ (right panel) on $R_f$. As a default, the length scale ratio (LSR) approach assumes $Pr_{t0} = 0.85$, $a_p$ = 0.33, and $c_P$ = 2.8. The sensitivities of the LSR-based predictions with respect to $a_p$ and $c_P$ coefficients are documented in both the panels. In addition, the predictions from the EFB approach are overlaid in these panels for comparison. The CSB-based result is also included in the left panel. Since the CSB and LSR approaches predict an identical relationship for normalized momentum flux, the CSB-based results are not shown in the right panel.}
\label{FigRatio2}
\end{figure*}

 As documented in Appendix~2, the CSB approach of Katul et al.~\cite{katul14} predicts: 
\begin{equation}
    R_{pw}^{CSB} = \left(\frac{c^{CSB}_T}{c^{CSB}_0}\right)\frac{R_f}{\left(1-R_f\right)},
    \label{CSB-Rpw1}
\end{equation}
where $c^{CSB}_0$ and $c^{CSB}_T$ equal to 0.65 and 0.80, respectively. On the other hand, according to the EFB approach of Zilitinkevich et al.~\cite{zilitinkevich13}, we have (refer to Appendix~3): 
\begin{equation}
    R^{EFB}_{pw} = \left(\frac{c^{EFB}_P}{A_z} \right)\frac{R_f}{(1-R_f)},
    \label{EFB-Rpw1}
\end{equation}
where, $c_P^{EFB}$ is 0.86. Zilitinkevich et al.~\cite{zilitinkevich13} assumed that the anisotropy parameter $A_z$ (discussed in the following section) varies from 0.2 (neutral condition) to 0.03 (strongly stratified condition). 

We intercompare Eqs.~(\ref{LSR-Rpw1}), (\ref{CSB-Rpw1}) and (\ref{EFB-Rpw1}) via Fig.~\ref{FigRatio2} (left panel). In comparison to the LSR approach, the CSB approach underestimates $R_{pw}$ by a factor of more than 2. The CSB approach makes an assumption that the temperature spectrum has a flat shape in the buoyancy range (refer to Appendix~2) which is not supported by field observations. We speculate that, as a consequence of this idealization, the CSB approach underestimates the variance of temperature, and in turn, underestimates $R_{pw}$. The predictions of the EFB approach and the LSR approach agree reasonably well up to $R_f \approx 0.15$. For higher stability conditions, the EFB predicts a sharp increase in $R_{pw}$ values. This drastic behavior can be attributed to the assumed stability-dependence of $A_z$ (see Fig.~6 of \cite{zilitinkevich13}).   

In the context of normalized momentum fluxes, the CSB and LSR approaches make identical predictions; please compare Eqs.~(\ref{LSR-Ruw}) and (\ref{CSB-Ruw}). However, the prediction from the EFB approach include terms involving $A_z$ in the numerator [refer to Eq.~(\ref{EFB-Ruw})]. Thus, owing to the assumed stability-dependence of $A_z$, the EFB approach predicts much higher value of normalized momentum fluxes in comparison to the LSR approach as depicted in the right panel of Fig.~\ref{FigRatio2}. Rigorous analyses of observational and simulated data will be needed to (in)validate these predictions.    

All the theoretical approaches predict an almost identical relationship for the normalized correlation of $w$ and $\theta$; refer to Eqs.~(\ref{LSR-RwT}), (\ref{CSB-RwT}), and (\ref{EFB-RwT}). The only difference arises due to the assumed value of $Pr_{t0}$. The LSR, CSB, and EFB approaches assume $Pr_{t0}$ to be equal to 0.85, 1, and 0.8, respectively.  

\section{Discussions}
\label{disc}

In this section, we elaborate on a few limitations of the proposed LSR approach and how to overcome them in a practical manner. 

\subsection{Vertical Anisotropy of Turbulence}
\label{Anisotropy}

In this study, we have used Eq.~(\ref{EqEDR2}) to parameterize energy dissipation rate ($\overline{\varepsilon}$). A more common practice would be to use Eq.~(\ref{EqEDRTKE}) or its following variant:   
\begin{subequations}
\begin{equation}
\overline{\varepsilon} = c_2^{*} \frac{q^2}{\left(\frac{L_X}{\sigma_w}\right)} = c_2^{*} \frac{\sigma_w^3}{A_z L_X},
\label{EqEDRTKE2}
\end{equation}
where,
\begin{equation}
    A_z = \frac{\overline{e}_w}{\overline{e}} = \frac{\sigma_w^2}{q^2},
    \label{EqAz1}
\end{equation}
\end{subequations}
and $c_2^{*}$ is an unknown coefficient. In Section~\ref{Derivations}, we have implicitly assumed $c_2^{*} A_z$ to be a constant ($c_2$). In the literature, there is some evidence that the anisotropy parameter, $A_z$, may be dependent on $Ri_g$. 

Based on observational and simulation data of turbulent air flows, Schumann and Gerz~\cite{schumann95} proposed the following empirical equation for $0 < Ri_g < 1$: 
\begin{equation}
    A_z = 0.15 + 0.02 Ri_g + 0.07 \exp{\left(-\frac{Ri_g}{0.25} \right)}.
    \label{EqAz2}
\end{equation}
According to this equation $A_z$ is weakly dependent on $Ri_g$; as a matter of fact, Schumann and Gerz~\cite{schumann95} stated ``the conclusions do not change much'' if $A_z = 0.22$ is used. Based on a DNS database, Basu et al.~\cite{basu21a} reported $A_z$ to be approximately equal to 0.18 for $0 < Ri_g < 0.2$. The parameterizations of Canuto et al.~\cite{canuto08}, Kantha and Clayson~\cite{kantha09}, and Cheng et al.~\cite{cheng20} predict gradual decrease of $A_z$ from near-neutral to strongly stratified conditions. Their predicted $A_z^{(Ri_g=0)}$ range from 0.22 to 0.26; whereas, $A_z^{(Ri_g > 1)}$ vary from about 0.15 to 0.20. In contrast, Zilitinkevich et al.~\cite{zilitinkevich13} used an empirical formulation which assumes $A_z^{(Ri_g=0)}$ = 0.20 and $A_z^{(Ri_g > 1)} \approx$ 0.03. The published datasets documented by Zilitinkevich et al.~\cite{zilitinkevich13} (see their Figure~6) and Cheng et al.~\cite{cheng20} (see their Figure~3c), in order to corroborate their respective formulations, do not portray any clear trends. A case in point are the wind tunnel measurements by Ohya~\cite{ohya01} which exhibit random fluctuating behavior. Surprisingly, a strongly increasing trend of $A_z$ with respect to $Ri_g$ was predicted by large-eddy simulation data of \cite{zilitinkevich07} (see their Figure 4); this was in direct contradiction to their analytical prediction. Given this diversity in the $A_z$-vs-$Ri_g$ relationship, we strongly recommend more research in this arena. 

If we utilize Eq.~(\ref{EqEDRTKE2}) instead of Eq.~(\ref{EqEDR2}), it is straightforward to re-derive all the equations reported in earlier sections. Some of the key equations are given here: 
\begin{widetext}
\begin{subequations}
\begin{equation}
    L_X = \sqrt{\frac{c_2^{*}}{c_1 A_z}} \left(\frac{\sigma_w}{S} \right) \left( \frac{1}{\sqrt{1-Ri_g/Pr_t}} \right),
    \label{EqAz3a}
\end{equation}
\begin{equation}
    \frac{L_H^2}{L_E^2} = \frac{\left(Pr_t - Ri_g \right)A_z}{c_P^{*}},
    \label{EqAz3b}
\end{equation}
\begin{equation}
    R_{pw} = \frac{c_P^{*} Ri_g}{\left(Pr_t - Ri_g\right) A_z} = \frac{c_P^{*} R_f}{\left(1-R_f\right) A_z}, 
    \label{EqAz3c}
\end{equation}
\begin{equation}
    \frac{R_{uw}}{R_{uw0}} = \sqrt{\frac{A_z^{(Ri_g=0)}}{A_z}}\left(\frac{1}{\sqrt{1 - Ri_g/Pr_t}}\right) = \sqrt{\frac{A_z^{(Ri_g=0)}}{A_z}}\left(\frac{1}{\sqrt{1-R_f}}\right).
    \label{EqAz3d}
\end{equation}
\label{EqAz3}
\end{subequations}
\end{widetext}
Here $c_P^{*}$ is an unknown coefficient and can be estimated following the procedure for $c_P$. Furthermore, the quadratic equation for the turbulent Prandtl number becomes: 
\begin{equation}
    Pr_t^2 - \left[c_5 + Ri_g + \frac{\left(1 - a_p\right) c_P^{*}}{A_z} Ri_g \right] Pr_t + c_5 Ri_g = 0.
    \label{EqAz4}
\end{equation}
                                           
\subsection{Imbalance of Production and Dissipation of TKE}

In Eq.~(\ref{EqEDR1}), we have assumed that the production and dissipation of TKE balances exactly. Following Schumann and Gerz~\cite{schumann95}, we can define their ratio, termed a `growth factor', as follows: 
\begin{equation}
    G  = \frac{-\left(\overline{u'w'}\right) S}{-\beta \overline{w'\theta'} + \overline{\varepsilon}}.
    \label{EqG1}
\end{equation}
It is likely that under strongly stratified condition, dissipation exceeds production. Thus, $G$ can become less than unity for high values of $Ri_g$. We can re-write Eq.~(\ref{EqG1}) as follows: 
\begin{equation}
    \overline{\varepsilon} = - \left(\overline{u'w'}\right) \frac{S}{G} + \beta \overline{w'\theta'} =  - \left(\overline{u'w'}\right) S^{*} + \beta \overline{w'\theta'}.
    \label{EqG2}
\end{equation}
The key equations will then become: 
\begin{subequations}
\begin{align}
    L_X &= c_H L_H \left( \frac{1}{\sqrt{1/G-Ri_g/Pr_t}} \right)\\ 
        &= c_H L_b \left( \frac{ \sqrt{Ri_g}}{\sqrt{1/G-Ri_g/Pr_t}} \right),
    \label{EqG3a}
\end{align}
\begin{equation}
    \frac{L_H^2}{L_E^2} = \frac{\left(Pr_t/G - Ri_g \right)}{c_P},
    \label{EqG3b}
\end{equation}
\begin{equation}
    R_{pw} = \frac{c_P Ri_g}{\left(Pr_t/G - Ri_g\right)} = \frac{c_P R_f}{\left(1/G-R_f\right)}, 
    \label{EqG3c}
\end{equation}
\begin{equation}
    \frac{R_{uw}}{R_{uw0}} = \frac{1}{\sqrt{1/G - Ri_g/Pr_t}} = \frac{1}{\sqrt{1/G-R_f}}.
    \label{EqG3d}
\end{equation}
\label{EqG3}
\end{subequations}
In this case, the quadratic equation for the turbulent Prandtl number becomes: 
\begin{equation}
    Pr_t^2 - \left[c_5 + Ri_g G + \left(1 - a_p\right) c_P G Ri_g \right] Pr_t + c_5 G Ri_g = 0.
    \label{EqG4}
\end{equation}

We would like to emphasize that the exact dependence of $G$ on stability is not well studied in the literature. Schumann and Gerz \cite{schumann95} proposed an empirical (exponential decay) equation for $G$-vs-$Ri_g$ based on limited data. We hypothesize that for very stable conditions ($Ri_g > 1$), $G$ should be proportional to $Ri_g^{-1}$. For practical applications, we propose the following heuristic parameterization for $G$:
\begin{equation}
    G = \min \left(1,Ri_g^{-1}\right).
    \label{LSR-G}
\end{equation}
Thus, for $Ri_g < 1$, $G$ equals to 1. In other words, production and dissipation of TKE balance each other for weakly and moderately stable condition. However, the balance is lost (i.e., $G < 1$) for very stable conditions.

If Eq.~(\ref{LSR-G}) is valid, then according to Eq.~(\ref{EqG3a}), $L_X$ will be approximately equal to the buoyancy length scale ($L_b$) for very stable conditions. Perhaps more interestingly, if Eq.~(\ref{LSR-G}) indeed holds, Eq.~(\ref{EqG4}) predicts that $Pr_t$ should saturate to a constant value for $Ri_g > 1$ . Such a prediction is not in agreement with some of the datasets reported in Fig.~(\ref{FigPr}). However, it is consistent with the findings reported by \cite{kitamura13} based on wind tunnel experiments and large-eddy simulations; refer to their Fig.~3.  

We would like to emphasize that Eq.~(\ref{LSR-G}) is based on a heuristic argument and has not been verified yet. In our future work, we will leverage on Eq.~(\ref{EqG3d}) to extract a reliable formulation for $G$.

\subsection{Combined Scenario}

For the most general case, one should account for the effects of both anisotropy and decay of TKE. In such a combined scenario, both $A_z$ and $G$ terms will appear in the aforementioned equations. For example, the length scale equation will read: \begin{equation}
    L_X = \sqrt{\frac{c_2^{*}}{c_1 A_z}} \left(\frac{\sigma_w}{S} \right) \left( \frac{1}{\sqrt{1/G-Ri_g/Pr_t}} \right).
\end{equation}
Similar to Eq.~(\ref{EqPrInfty}), for very stable condition (i.e., $Ri_g \gg 1$), the Prandtl number equation will be simplified to: 
\begin{equation}
    Pr_t \approx \left(G + \frac{\left(1 - a_p\right) c_P^{*} G}{A_z} \right) Ri_g = \frac{Ri_g}{R_{f\infty}},
\end{equation}
Thus, the exact value of $R_{f\infty}$ depends on $a_p$, $A_z$, $G$ and $c_P^{*}$. Since stability dependencies of $a_p$, $A_z$ and $G$ are rather uncertain, empirical parameterizations for the combined terms (e.g., $G/A_z$) might be more practical for certain applications. High quality data from laboratory experiment (e.g., wind tunnel) and/or direct numerical simulation will be needed to derive such parameterizations. 

\section{Conclusions}

In this study, we have analytically derived an explicit relationship between the Prandtl number and the gradient Richardson number. Our derivation is rather simple from a mathematical standpoint and does not make elaborate assumptions beyond variance and sensible heat flux budget equations. Most of the unknown coefficients of the proposed relationship are easily estimated from well-known surface layer similarity relationships. Our proposed Prandtl number formulation agrees very well with other competing approaches of quite different theoretical foundations and assumptions.  

Our original analysis can be easily extended to include the effects of vertical anisotropy. It can also account for an imbalance of production and dissipation of TKE under very stable conditions. We have provided generalized formulations to account for these effects. However, these generalized formulations require stability-dependent formulations for a few parameters (e.g., $A_z$, $G$) which are not well established in the literature. Currently, we are analyzing wind tunnel measurements and DNS-generated datasets to derive these formulations in a robust manner.  

One of the limitations of the present study is that, for simplicity, it omits any discussion of internal gravity waves \cite{staquet02,sun15}. However, in stable boundary layers, specially under strong stratification, wave-turbulence interactions are extremely important. Thus far, only a handful of analytical studies have looked into such interactions \cite{kleeorin19,kurbatskii2019,sukoriansky08,zilitinkevich09}. We hope to further advance our proposed LSR approach along this direction in the future.   

\section*{Acknowledgements}
We are truly grateful to Hubert Luce for independently cross-checking our analytical derivations and in the process detecting a bug in one of the coefficients. The first author is indebted to Gabriel Katul and Dan Li for in-depth scientific exchanges on the co-spectral budget formulation and for confirming our derivations in Appendix~2. We also thank Lakshmi Kantha and Margaret Lemone for providing constructive feedback.

\bibliographystyle{spbasic_updated}      
\bibliography{OLS}   

\section*{Appendix 1: Parameterization of the Pressure-Temperature Interaction Term}
\label{Appendix_rotta}

In the prognostic equation of sensible heat flux ($\overline{u'_i \theta'}$), a pressure-temperature interaction term $\Pi_i = \left(-\frac{1}{\rho_0}\overline{\theta' \frac{\partial p'}{\partial x_i}}\right)$ appears \cite{arya75,garratt92}. This loss term is significant for atmospheric boundary layer (ABL) flows and requires a reliable parameterization. Using the product rule of calculus, $\Pi_i$ can be decomposed as follows \cite{hanjalic11,kantha94,stull88}: 
\begin{equation}
    -\frac{1}{\rho_0}\overline{\theta' \frac{\partial p'}{\partial x_i}} = - \frac{\partial}{\partial x_k} \left(\frac{1}{\rho_0} \overline{p'\theta'} \delta_{ik} \right) + \frac{1}{\rho_0} \overline{p' \frac{\partial \theta'}{\partial x_i}}.
    \label{rotta1}
\end{equation}
Here $p$ and $\rho_0$ denote pressure and a reference density, respectively. The symbol $\delta_{ik}$ represents Kronecker delta. 

The first term on the right hand side of Eq.~(\ref{rotta1}) represents turbulent diffusion of temperature field by pressure fluctuations and is sometimes neglected under the assumption of isotropy or using scaling argument \cite{stull88}. As an alternative, in a number of modeling studies, it has been combined with the turbulent transport term, and in turn, the total term is parameterized via K-theory \cite{moeng86}.   

The second term $\left(\Phi_i = \frac{1}{\rho} \overline{p' \frac{\partial \theta'}{\partial x_i}}\right)$ is known as the pressure scrambling of the fluctuating temperature field. This term is split into three separate components representing different interactions \cite{cheng02,hanjalic11}: 
\begin{equation}
    \Phi_i = \Phi_i^{TT} + \Phi_i^{S} + \Phi_i^{B}.
    \label{rotta11}
\end{equation}
The term $\Phi_i^{TT}$ captures turbulence-turbulence interactions. Following Rotta's celebrated return-to-isotropy hypothesis \cite{rotta51}, Monin \cite{monin65} parameterized this term as follows: 
\begin{equation}
    \Phi_i^{TT} = - \frac{\overline{u'_i \theta'}}{\tau_R},
    \label{rotta12}
\end{equation}
where $\tau_R$ is the return-to-isotropy time scale. In the absence of external forces, this term relaxes turbulence to an isotropic state with zero overall heat flux \cite{umlauf05}. Even though Eq.~(\ref{rotta12}) is the most popular in the literature, alternative parameterizations for $\Phi_i^{TT}$ have been proposed in the past (please refer to \cite{hanjalic11}).  

The mean shear-turbulence interaction is denoted by $\Phi_i^{S}$ and is parameterized as follows \cite{andren93,chen98,hanjalic11}:
\begin{equation}
    \Phi_i^{S} = a_s \overline{u'_k \theta'} \frac{\partial \overline{u}_i}{\partial x_k}. 
    \label{rotta13}
\end{equation}
In the absence of significant subsidence or under quiescent synoptic condition, the vertical component (i.e., $\Phi_3^{S}$) is negligible in the ABL flows since $\overline{u}_3 = \overline{w} \approx 0$; a comprehensive modeling study by \cite{andren93} provides supporting results. 

The following equation is often used for representing the buoyancy-turbulence interaction \cite{hanjalic11,launder75}:
\begin{equation}
    \Phi_i^{B} = -a_p \beta \sigma_\theta^2 \delta_{i3}.  
    \label{EqPhiB}
\end{equation}
Even though this term is known to be important for non-neutral flows, quite interestingly, the well-known parameterizations of Mellor and Yamada \cite{mellor82} disregarded it. 

\begin{table}[t]
\caption{Recommended values of $a_s$ and $a_p$ coefficients}
\label{T2}       
\begin{tabular}{lll}
\noalign{\smallskip}\hline
Study & $a_s$ & $a_p$\\
\noalign{\smallskip}\hline
Launder \cite{launder75} & 0.50 & 0.50\\
Moeng and Wyngaard \cite{moeng86} & -- & 0.50\\
Andr\'{e}n and Moeng \cite{andren93} & 0.75 & --\\
Kantha and Clayson \cite{kantha94} & 0.70 & 0.20\\
Nakanishi \cite{nakanishi01} & 0.65 & 0.294 \\
\hline\noalign{\smallskip}
\end{tabular}
\end{table}

Over the years, various studies recommended different sets of values for $a_s$ and $a_p$. Some of them are documented in Table~\ref{T2}. Additionally, an empirical stability-dependent formulation for $a_p$ was proposed by Wyngaard \cite{wyngaard75}:
\begin{subequations}
\begin{equation}
    a_p = 0.5 + 1.5 Ri_g^2 - Ri_g^3 \mbox{\hspace{0.1in} for \hspace{0.1in}} 0 < Ri_g < 1
    \label{ap1}
\end{equation}
\begin{equation}
    a_p = 1 \mbox{\hspace{0.1in} for \hspace{0.1in}} Ri_g > 1.
    \label{ap2}
\end{equation}
\end{subequations}
However, $a_p$ = 1 for $Ri_g > 1$ does not lead to a physically meaningful solution when used in conjunction with Eq.~(\ref{PrTFinal}). It is trivial to show that the solutions of the quadratic equation lead to two solutions: (i) $Pr_t = Pr_{t0}$ and (ii) $Pr_t$ = $Ri_g$. Neither of these solutions are plausible for the strongly stratified regime. In lieu of a realistic stability-dependent parameterization, in this study, we have decided to set $a_p$ as a fixed coefficient and have performed simple sensitivity analysis to quantify its influence on the overall predictions.  

By combining Eqs.~(\ref{rotta1}--\ref{rotta13}), the overall pressure-temperature interaction term for the vertical component of sensible heat flux can be simplified as follows: 
\begin{equation}
    -\frac{1}{\rho_0}\overline{\theta' \frac{\partial p'}{\partial z}} = - \frac{\overline{w' \theta'}}{\tau_R} -a_p \beta \sigma_\theta^2.
    \label{rotta14}
\end{equation}
The terms on the right hand side of Eq.~(\ref{rotta14}) are included in the simplified budget equation [i.e., Eq.~(\ref{EqCHIW1})] for sensible heat flux. The other terms of Eq.~(\ref{EqCHIW1}) account for productions due to mean gradient ($-\sigma_w^2 \Gamma$) and buoyancy ($\beta \sigma^2_\theta$). 

\section*{Appendix 2: Co-spectral Budget (CSB) Approach}
\label{Appendix_csb}

In this section, we re-derive the relevant equations of the co-spectral budget (CSB) approach following the footsteps of Katul et al.~\cite{katul14}. Along the way, we point out some of their (implicit) assumptions and differences to our newly proposed LSR approach. During this exercise, we noted certain sign errors in the original derivations of \cite{katul14}. D.~Li~\cite{li21} confirmed our findings and pointed out additional sign errors in \cite{katul14}. Fortunately, all these errors cancel out and do not have any effect on the key results. We have communicated our findings to G.~G.~Katul~\cite{katul21} and he has kindly verified them. 

The starting point of the CSB approach is vertical sensible heat and momentum flux budget equations in wavenumber space: 
\begin{subequations}
\begin{align}
    \underbrace{P_{w\theta}(k_x)}_\text{production} + \underbrace{\beta F_{\theta\theta}(k_x)}_\text{buoyancy} + \underbrace{\Pi_\theta(k_x)}_{\substack{\text{pressure-temperature}\\ \text{decorrelation}}} & = 0,
    \label{CSB-Budget}\\
    \underbrace{P_{uw}(k_x)}_\text{production} + \underbrace{\Pi_u(k_x)}_{\substack{\text{pressure-velocity}\\ \text{decorrelation}}} & = 0.
    \label{CSB-Budget2}
\end{align}
\end{subequations}
Here $F_{\theta\theta}$ is the one dimensional temperature spectrum. $k_x$ denotes wavenumber in the along-wind direction. Both these equations assume steady-state condition. They neglect turbulent transport and molecular diffusion terms. Interestingly, the momentum flux equation also neglects the buoyancy term. 

The pressure-temperature and pressure-velocity interactions are parameterized as follows: 
\begin{subequations}
\begin{align}
    \Pi_\theta(k_x) & = -A_T \frac{F_{w\theta}}{\tau(k_x)} - c^{CSB}_{1T} P_{w\theta}(k_x),
    \label{CSB-Rotta}\\
    \Pi_u(k_x) & = -A_u \frac{F_{uw}}{\tau(k_x)} - c^{CSB}_{1U} P_{uw}(k_x).
    \label{CSB-Rotta2}
\end{align}
\end{subequations}
Where $F_{w\theta}$ and $F_{uw}$ are the cospectra between $w$--$\theta$ and $w$--$u$, respectively. $\tau(k_x)$ is a relaxation time-scale. $A_T$, $A_U$, $C^{CSB}_{1T}$, and $C^{CSB}_{1U}$ are constants which should be prescribed. Katul et al.~\cite{katul14} assumed: $A_T = A_U = 1.8$,  and $c^{CSB}_{1T} = c^{CSB}_{1U} = 3/5$. 

Please note that Eq.~(\ref{CSB-Rotta}) does not include the commonly used buoyancy-turbulence interaction term [see Eq.~(\ref{EqPhiB}) in Appendix~1]. Instead, it includes an unorthodox term which is proportional to the production term. 

The production terms are expressed as follows:
\begin{subequations}
\begin{align}
    P_{w\theta}(k_x) & = -\Gamma F_{ww}(k_x),
    \label{CSB-Production}\\
    P_{uw}(k_x) & = -S F_{ww}(k_x).
    \label{CSB-Production2}
\end{align}
\end{subequations}
Here, the one dimensional vertical velocity spectrum is denoted by $F_{ww}$. Please note that both these equations in \cite{katul14} contain sign errors as pointed out by \cite{li21}. 

By combining Eqs.~(\ref{CSB-Budget}), (\ref{CSB-Rotta}), and (\ref{CSB-Production}), we get: 
\begin{subequations}
\begin{equation}
    F_{w\theta} = -\left(\frac{\tau(k_x)}{A_T}\right) \left[\left(1-c^{CSB}_{1T} \right) \Gamma F_{ww}(k_x) - \beta F_{\theta\theta}(k_x) \right].
    \label{CSB-FwT1}
\end{equation}
Similarly, by using Eqs.~(\ref{CSB-Budget2}), (\ref{CSB-Rotta2}), and (\ref{CSB-Production2}), we arrive at: 
\begin{equation}
F_{uw} = -\left(\frac{\tau(k_x)}{A_U}\right) \left[\left(1-c^{CSB}_{1U} \right) S F_{ww}(k_x) \right].
    \label{CSB-Fuw1}
\end{equation}
\end{subequations}

Next, Katul et al.~\cite{katul14} assumed that $F_{ww}(k_x)$, $F_{\theta\theta}(k_x)$, and $\tau(k_x)$ follow the inertial-range scaling behavior within the range $k_a \le k \le \infty$ as hypothesized by Kolmogorov-Obukhov-Corrsin: 
\begin{subequations}
\begin{align}
    F_{ww}(k_x) & = c^{CSB}_0 \overline{\varepsilon}^{2/3} k_x^{-5/3},
    \label{CSB-Fww1}\\
    F_{\theta}(k_x) & = c^{CSB}_T \left(\overline{\varepsilon}\right)^{-1/3} \overline{N}_\theta k_x^{-5/3},
    \label{CSB-FTT1}\\    
    \tau(k_x)   & = \left(\overline{\varepsilon}\right)^{-1/3} k_x^{-2/3}.
\end{align}
\label{CSB-KOC}
\end{subequations}
Where $\overline{N}_\theta$ is simply half of dissipation rate of temperature variance ($\overline{\chi}_\theta$). $c^{CSB}_0$ and $c^{CSB}_T$ are supposed to be universal constants. Katul et al.~\cite{katul14} assumed $c^{CSB}_0 = 0.65$ and $c^{CSB}_T = 0.80$. Please note that, for simplicity, they assumed that the inertial-range scaling also holds in the dissipation range. 

For low wavenumbers ($0 \le k_x \le k_a$), the CSB approach assumes flat (i.e., white noise) spectra: 
\begin{subequations}
\begin{align}
    F_{ww}(k_x) & = c^{CSB}_0 \overline{\varepsilon}^{2/3} k_a^{-5/3},
    \label{CSB-Fww2}\\
    F_{\theta\theta}(k_x) & = c^{CSB}_T \left(\overline{\varepsilon}\right)^{-1/3} \overline{N}_\theta k_a^{-5/3},
    \label{CSB-FTT2}\\
    \tau(k_x) & = \left(\overline{\varepsilon}\right)^{-1/3} k_a^{-2/3}.
\end{align}
\label{CSB-Buoyancy}
\end{subequations}
Over the decades, several competing hypotheses (e.g., \cite{bolgiano59,bolgiano62,lumley64b,monin65b,shur62}) have been put forward to characterize the low wavenumber (aka buoyancy-range) spectra. We would like to point out that none of these hypotheses are in line with the assumption of the CSB approach. Furthermore, in the surface layer, there are ample evidence in the literature (e.g., \cite{kader91,li16b}) that temperature spectra follow $k_x^{-1}$ scaling and not $k_x^0$ scaling as assumed by the CSB approach.   

By integrating and summing Eqs.~(\ref{CSB-Fww1}) and (\ref{CSB-Fww2}) we get:
\begin{equation}
\begin{split}
    \sigma_w^2 & = \int_0^{k_a} F_{ww}(k_x) dk_x + \int_{k_a}^{\infty} F_{ww}(k_x) dk_x\\
    & = \int_0^{k_a} c^{CSB}_0 \overline{\varepsilon}^{2/3} k_a^{-5/3} dk_x + \int_{k_a}^{\infty} c^{CSB}_0 \overline{\varepsilon}^{2/3} k_x^{-5/3} dk_x\\
    & = \frac{5}{2} c^{CSB}_0 \overline{\varepsilon}^{2/3} k_a^{-2/3}.
    \label{CSB-varW}
\end{split}
\end{equation}
Similarly, from Eqs.~(\ref{CSB-FTT1}) and (\ref{CSB-FTT2}) we get:
\begin{widetext}
\begin{equation}
\begin{split}
    \sigma_\theta^2 & = \int_0^{k_a} F_{\theta\theta}(k_x) dk_x + \int_{k_a}^{\infty} F_{\theta\theta}(k_x) dk_x\\
    & = \int_0^{k_a} c^{CSB}_T \left(\overline{\varepsilon}\right)^{-1/3} \overline{N}_\theta k_a^{-5/3} dk_x + \int_{k_a}^{\infty} c^{CSB}_T \left(\overline{\varepsilon}\right)^{-1/3} \overline{N}_\theta k_x^{-5/3} dk_x\\
    & = \frac{5}{2} c^{CSB}_T \left(\overline{\varepsilon}\right)^{-1/3} \overline{N}_\theta k_a^{-2/3}.
    \label{CSB-varT}
\end{split}
\end{equation}
\end{widetext}
By making use of Eq.~(\ref{CSB-FwT1}) in conjunction with Eqs.~(\ref{CSB-KOC})--(\ref{CSB-Buoyancy}), it is straightforward to derive:   
\begin{subequations}
\begin{equation}
\begin{split}
\overline{w'\theta'} & = \left[ \int_0^{k_a} F_{w\theta}(k_x) dk_x + \int_{k_a}^{\infty} F_{w\theta}(k_x) dk_x \right],\\
                    & = - \left(\frac{7 c^{CSB}_0 \Gamma \left(\overline{\varepsilon}\right)^{1/3} Q}{10 A_T}\right) k_a^{-4/3},
\end{split}
\label{CSB-wT}
\end{equation}
where 
\begin{equation}
Q = \left[1 - \frac{\beta c^{CSB}_T \overline{N}_\theta}{(1 - c^{CSB}_{1T}) c^{CSB}_0 \Gamma \overline{\varepsilon}} \right].
\label{CSB-Q}
\end{equation}
\end{subequations}
Both these equations in \cite{katul14} contain sign errors.  

In an analogous manner, we get from Eq.~(\ref{CSB-Fuw1}) and Eqs.~(\ref{CSB-KOC})--(\ref{CSB-Buoyancy}): 
\begin{equation}
\begin{split}
\overline{u'w'} & = \left[ \int_0^{k_a} F_{uw}(k_x) dk_x + \int_{k_a}^{\infty} F_{uw}(k_x) dk_x \right],\\
                & = - \left(\frac{7 c^{CSB}_0 S \left(\overline{\varepsilon}\right)^{1/3}}{10 A_U}\right) k_a^{-4/3}.
\end{split}
\label{CSB-uw}
\end{equation}
This equation in \cite{katul14} contains a sign error.  

From Eqs.~(\ref{CSB-wT}), (\ref{CSB-Q}), and (\ref{CSB-uw}), we have: 
\begin{equation}
    Pr_t = \frac{-\overline{u'w'}/S}{-\overline{w'\theta'}/\Gamma} = \frac{A_T}{A_U Q} = \frac{1}{Q}.
\end{equation}
Note that Katul et al.~\cite{katul14} assumed $A_T = A_U$. 

The budget equations of TKE and $\sigma_\theta^2$ can be written as: 
\begin{subequations}
\begin{align}
    \overline{\varepsilon} & = - \left(\overline{u'w'}\right) S + \beta \overline{w'\theta'},
    \label{CSB-EqEDR1}\\
    \overline{N}_{\theta} & = - \left(\overline{w'\theta'}\right) \Gamma,
    \label{CSB-EqCHI1}
\end{align}
\end{subequations}
Unfortunately, a sign error appears in the equation for $\overline{N}_\theta$ in Katul et al.~\cite{katul14}. 

Dividing Eq.~(\ref{CSB-EqCHI1}) by Eq.~(\ref{CSB-EqEDR1}) and using the definition of flux Richardson number ($R_f$), we can write: 
\begin{equation}
    \left(\frac{\beta \overline{N}_\theta}{\Gamma \overline{\varepsilon}}\right) = \frac{R_f}{1-R_f}.
    \label{CSB-XXX}
\end{equation}
Hence, 
\begin{subequations}
\begin{equation}
    Q = 1 - \frac{c^{CSB}_T}{\left(1-c^{CSB}_{1T}\right)c^{CSB}_0} \left(\frac{R_f}{1-R_f}\right) = \left(\frac{1 - \omega^{CSB}                     R_f}{1-R_f}\right),
\end{equation}
and 
\begin{equation}
    Pr_t = \left(\frac{1-R_f}{1-\omega^{CSB} R_f}\right).
    \label{CSB-Pr}
\end{equation}
Where,
\begin{equation}
\omega^{CSB} = 1 + \frac{c^{CSB}_T}{\left(1-c^{CSB}_{1T}\right)c^{CSB}_0}.     
\end{equation}
\end{subequations}
Katul et al.~\cite{katul14} assumed $c^{CSB}_0$, $c^{CSB}_T$ and $c^{CSB}_{1T}$ to be equal to $0.65$, $0.80$, and $3/5$, respectively. As a result, $\omega^{CSB} \approx 4$.

From Eq.~(\ref{CSB-Pr}), we can easily derive the following quadratic equation (not reported in previous CSB-related publications): 
\begin{subequations}
\begin{equation}
    Pr_t^2 - \left(1 + \omega^{CSB} Ri_g \right) Pr_t + Ri_g = 0,
\end{equation}
and its roots are: 
\begin{equation}
    Pr_t = \frac{\left(1 + \omega^{CSB} Ri_g \right) \pm \sqrt{\left(1 + \omega^{CSB} Ri_g \right)^2 - 4Ri_g}}{2}.
    \label{CSB-PrTFinal}
\end{equation}
\end{subequations}
Only the larger root is physically meaningful.  

We would like to point out that Eq.~(\ref{CSB-PrTFinal}) is cast in a different analytical form than the original CSB formulation in \cite{katul14} and follow-up studies. For neutral condition ($Ri_g = 0$), according to Eq.~(\ref{CSB-PrTFinal}), $Pr_{t0}$ equals to 1. Whereas, according Eq.~(37) of \cite{katul14}, $Pr_{t0}$ is undetermined for neutral condition. 

Using Eqs.~(\ref{CSB-varW}), (\ref{CSB-varT}), and (\ref{CSB-XXX}), the ratio of turbulent potential and kinetic energies can be derived as follows: 
\begin{equation}
\begin{split}
R^{CSB}_{pw} & = \left( \frac{\beta}{N}\right)^2 \frac{\sigma_\theta^2}{\sigma_w^2} \\  
       & = \left(\frac{\beta}{\Gamma} \right) \left( \frac{c^{CSB}_T \overline{N}_\theta}{c^{CSB}_0 \overline{\varepsilon}}\right)\\
       & = \frac{c^{CSB}_T}{c^{CSB}_0} \left(\frac{R_f}{1-R_f}\right).
\end{split}
\end{equation}

For neutral condition, Eq.~(\ref{CSB-EqEDR1}) simplifies to: 
\begin{equation}
    \overline{\varepsilon}_0 = -\left(\overline{u'w'}_0\right) S_0
\end{equation}
Thus, 
\begin{equation}
    \frac{\overline{u'w'}}{\overline{u'w'}_0} = \left(\frac{\overline{\varepsilon} S_0}{\overline{\varepsilon}_0 S}\right) \left(\frac{1}{1-R_f}\right).
\end{equation}
Utilizing this equation in conjunction with Eqs.~(\ref{CSB-varW}) and (\ref{CSB-uw}), after a little algebraic manipulation, we can derive the ratio of normalized momentum flux as: 
\begin{equation}
    \left(\frac{R_{uw}}{R_{uw0}}\right)^{CSB} = \frac{1}{\sqrt{1-R_f}}.
    \label{CSB-Ruw}
\end{equation}
This equation is identical to the prediction by the LSR approach [see Eq.~(\ref{LSR-Ruw})].  

Using Eqs.~(\ref{CSB-varW}), (\ref{CSB-varT}), (\ref{CSB-wT}), (\ref{CSB-Q}), and (\ref{CSB-EqCHI1}), we can deduce an expression for the normalized correlation between vertical velocity and potential temperature as follows: 
\begin{equation}
    \left(\frac{R_{w\theta}}{R_{w\theta 0}}\right)^{CSB} = \sqrt{\frac{Q}{Q_0}} = \sqrt{\frac{Pr_{t0}}{Pr_t}} = \frac{1}{\sqrt{Pr_t}}.
    \label{CSB-RwT}
\end{equation}
For neutral condition, by definition $Q$ equals to one. Thus, $Pr_{t0}$ is also unity.  

\section*{Appendix 3: Energy-and Flux-Budget (EFB) Approach}
\label{Appendix_efb}

Over the past several years, Zilitinkevich and his co-workers have proposed the so-called energy-and flux-budget (EFB) approach and its several modifications. In this appendix, we briefly discuss some of the salient features of this approach. We follow one of the later versions of the EFB approach as documented by Zilitinkevich et al.~\cite{zilitinkevich13}. 

The EFB approach makes use of the steady-state budget equations for both sensible heat and momentum fluxes. As a reminder to the readers, our proposed LSR approach does not utilize the momentum flux equation. In the case of the sensible heat flux equation, \cite{zilitinkevich13} parameterizes the pressure-temperature interaction term as follows: 
\begin{subequations}
\begin{equation}
    \frac{1}{\rho_0} \overline{\theta'\frac{\partial p'}{\partial z}} = \left(1 - c^{EFB}_\theta \right) \beta \sigma_\theta^2, 
\end{equation}
where $c^{EFB}_\theta$ is an unknown coefficient. In the LSR approach, we use the term $a_p$ to denote $(1-c^{EFB}_\theta)$. Interestingly, Zilitinkevich et al.~\cite{zilitinkevich13} neglects the commonly used turbulence-turbulence interactions [i.e., Eq.~(\ref{rotta12})] in the pressure-temperature interaction term. However, they use this exact term to parameterize the dissipation term (commonly neglected in the literature) of the sensible heat flux equation as follows:
\begin{equation}
    \varepsilon_z^{(F)} = \frac{\overline{w'\theta'}}{c^{EFB}_F \tau_\varepsilon}.
\end{equation}
Where, $\tau_\varepsilon$ is the dissipation time scale and $c^{EFB}_F$ is an unknown coefficient, assumed to be equal to 0.25. 
\end{subequations}
Effectively, both the EFB and the LSR approaches use the same form of parameterized sensible heat flux equation. From this equation, with minor algebraic manipulations, \cite{zilitinkevich13} derived:
\begin{subequations}
\begin{equation}
\overline{w'\theta'} = -K_H \Gamma = - 2 c^{EFB}_F \tau_\varepsilon \left(\overline{e}_w - c^{EFB}_\theta \overline{e}_p\right) \Gamma,    
\end{equation}
or,
\begin{equation}
    K_H = 2 c^{EFB}_F \tau_\varepsilon \left(\overline{e}_w - c^{EFB}_\theta \overline{e}_p\right).
    \label{efb-Kh}
\end{equation}
\end{subequations}
Please refer to Eqs.~(\ref{ep}) and (\ref{ew}) for the definitions of $\overline{e}_p$ and $\overline{e}_w$, respectively. 

In the case of the momentum flux equation, Zilitinkevich et al.~\cite{zilitinkevich13} makes several approximations. They neglect the dissipation term. In addition, they combine the buoyancy and pressure-velocity interaction terms and call it an `effective dissipation rate'. This combined term is parameterized like a return-to-isotropy term. The resultant momentum flux equation is written as follows: 
\begin{subequations}
\begin{equation}
    \overline{u'w'} = - K_M S = - 2 c^{EFB}_\tau \tau_\varepsilon \overline{e}_w S,
\end{equation}
where, $c^{EFB}_\tau$ is an unknown coefficient, assumed to be equal to 0.2. Thus, the eddy diffusivity can be represented as: 
\begin{equation}
    K_M = 2 c^{EFB}_\tau \tau_\varepsilon \overline{e}_w.
    \label{efb-Km}
\end{equation}
\end{subequations}
Based on Eqs.~(\ref{efb-Kh}) and (\ref{efb-Km}), one can write:  
\begin{equation}
    Pr_t = \frac{K_M}{K_H} =  \frac{\left(\frac{c^{EFB}_\tau}{c^{EFB}_F}\right)}{\left(1 - c^{EFB}_\theta \frac{\overline{e}_p}{\overline{e}_w} \right)}.
    \label{EFB-Pr}
\end{equation}
Zilitinkevich et al.~\cite{zilitinkevich13} argued that if $Pr_t \to \infty$ as $Ri_g \to \infty$, then in the limiting case: 
\begin{equation}
c^{EFB}_\theta = \left(\frac{\overline{e}_w}{\overline{e}_p} \right)_{Ri_g\to\infty}.    
\end{equation}
Even though this equation is only valid for $Ri_g \to \infty$, the EFB approach uses $c^{EFB}_\theta$ as a constant, being equal to 0.105, for all stability conditions. In our proposed LSR approach, the related coefficient is $(1- a_p)$ and we have also assumed it to be a constant in lieu of a reliable stability-dependent parameterization. 

From the budget equations of TKE and variance of potential temperature, along with the definition of flux Richardson number ($R_f$), \cite{zilitinkevich13} derived the following ratios: 
\begin{subequations}
\begin{equation}
\frac{\overline{e}}{\overline{e} + \overline{e}_p} = \frac{1-R_f}{1 - \left(1 - c^{EFB}_P\right) R_f},     
\end{equation}
and,
\begin{equation}
\frac{\overline{e}_p}{\overline{e} + \overline{e}_p} = \frac{c^{EFB}_P R_f}{1 - \left(1 - c^{EFB}_P\right) R_f}.        
\label{EFB-ep}
\end{equation}
\end{subequations}
Where $c^{EFB}_P$ is an unknown coefficient. Based on available data, \cite{zilitinkevich13} assumed it to be equal to 0.86.

By plugging in Eq.~(\ref{EFB-ep}) in Eq.~(\ref{EFB-Pr}) and using the definition $A_z = \frac{\overline{e}_w}{\overline{e}}$, one gets the following equation after simplification: 
\begin{equation}
   Pr_t = \frac{\left(\frac{c^{EFB}_\tau}{c^{EFB}_F}\right)}{\left(1 - c^{EFB}_\theta c^{EFB}_P \frac{R_f}{A_z (1-R_f)}\right)}. 
   \label{EFB-Pr2}
\end{equation}
For neutral condition (i.e., $R_f = 0$), with the chosen values of $c^{EFB}_\tau$ and $c^{EFB}_F$, the EFB approach predicts $Pr_{t0} = 0.8$. 

Please note that Eq.~(\ref{EFB-Pr2}) requires a parameterization for $A_z$. Zilitinkevich et al.~\cite{zilitinkevich13} proposed heuristic equations for the redistribution of TKE among various velocity components due to the effects of stratification. Those equations lead to a specific formulation for $A_z$; please refer to Eq.~(50c) of \cite{zilitinkevich13}. Using limited data, they further assumed $A_z^{(Ri_g = 0)} = 0.2$ and $A_z^{(Ri_g \to \infty)} = 0.03$. In Section~\ref{Anisotropy} we have provided more information on $A_z$. 

It is straightforward to derive the following normalized variances and fluxes from the EFB approach (see \cite{li16}): 
\begin{subequations}
\begin{equation}
R^{EFB}_{pw} = \frac{c^{EFB}_P R_f}{A_z(1-R_f)},
\label{EFB-Rpw}
\end{equation} 
\vspace{0.1in}
\begin{equation}
\left(\frac{R_{uw}}{R_{uw0}}\right)^{EFB} = \frac{\sqrt{\frac{A_z^{(Ri_g=0)}}{A_z}}}{\sqrt{1-R_f}},    
\label{EFB-Ruw}
\end{equation}
\begin{equation}
\left(\frac{R_{w\theta}}{R_{w\theta 0}}\right)^{EFB} = \sqrt{\frac{Pr_{t0}}{Pr_t}}.
\label{EFB-RwT}
\end{equation}
\end{subequations}
In Section~\ref{Comparison}, we have compared these equations against the predictions from the LSR and the CSB approaches. 

\end{document}